\documentclass[11pt]{article}
\pdfoutput=1
\usepackage{amsfonts,amsmath,amssymb,color,float}
\usepackage{graphicx}

\newfont{\twelvemsb}{msbm10 scaled\magstep1}
\newfont{\eightmsb}{msbm8}
\newfam\msbfam
\textfont\msbfam=\twelvemsb \scriptfont\msbfam=\eightmsb
\catcode`\@=11
\def\Bbb{\ifmmode\let\next\Bbb@\else
\def\next{\errmessage{Use \string\Bbb\space only in math mode}}\fi\next}
\def\Bbb@#1{{\fam\msbfam{{#1}}}}

\newcommand{\be}{\begin{equation}}
\newcommand{\ee}{\end{equation}}
\newcommand{\ba}{\begin{eqnarray}}
\newcommand{\ea}{\end{eqnarray}}

\begin{document}
\sloppy
\renewcommand{\thefootnote}{\fnsymbol{footnote}}
\newpage
\setcounter{page}{1} \vspace{0.7cm}
\begin{flushright}
NORDITA 2020-035
\end{flushright}
\vspace*{1cm}
\begin{center}
{\bf  $QQ$-system and non-linear integral equations for scattering amplitudes at strong coupling}\\
\vspace{1.8cm} {\large Davide Fioravanti $^a$, Marco Rossi $^b$, Hongfei Shu $^{c}$
\footnote{E-mail: fioravanti@bo.infn.it, rossi@cs.infn.it, hongfei.shu@su.se}}\\
\vspace{.5cm} $^a${\em Sezione INFN di Bologna, Dipartimento di Fisica e Astronomia,
Universit\`a di Bologna\\
Via Irnerio 46, 40126 Bologna, Italy} \\
\vspace{.3cm} $^b${\em Dipartimento di Fisica dell'Universit\`a della Calabria and
INFN, Gruppo collegato di Cosenza\\
Arcavacata di Rende, 87036 Cosenza, Italy} \\
\vspace{.3cm} $^c${\em Nordita, KTH Royal Institute of Technology and Stockholm University \\
Roslagstullsbacken 23, SE-106 91 Stockholm, Sweden}\\
\end{center}
\renewcommand{\thefootnote}{\arabic{footnote}}
\setcounter{footnote}{0}
\begin{abstract}
We provide the two fundamental sets of functional relations which describe the strong coupling limit of scattering amplitudes in $\mathcal{N} = 4$ SYM dual to Wilson loops in $AdS_3$: the basic $QQ$-system and the derived $TQ$-system. We use the $TQ$ relations and the knowledge of the main properties of the $Q$-function (eigenvalue of some $Q$-operator) to write the Bethe Ansatz equations, {\it viz.} a set of ('complex') non-linear-integral equations, whose solutions give exact values to the strong coupling amplitudes/Wilson loops. Moreover, they have some advantages with respect to the ('real') non-linear-integral equations of Thermodynamic Bethe Ansatz and still reproduce, both analytically and numerically,  the findings coming from the latter. In any case, these new functional and integral equations give a larger perspective on the topic. 

\end{abstract}
\vspace{1cm} 
\newpage

\section{Some background and plan of the paper}

Maximally supersymmetric $\mathcal{N} = 4$ gauge theory in 4D has been revealing many integrable features, mainly in the planar limit, which also involve its strong coupling dual string theory \cite {MGKPW}: for instance, integrable spin-chain hamiltonians and $S$-matrices appear in the computation of anomalous dimensions of composite operators ({\it cf. e.g.} \cite{BEI} and \cite{TBA-AdS-CFT} for the related Thermodynamic Bethe Ansatz to be compared with the following). Furthermore, a gluon scattering amplitude equals its dual, namely the null polygonal Wilson loop (WL)  and the latter admits an expression as quantum string partition function \cite{AM-amp}. In particular, at strong coupling, the amplitude is dominated by the classical contribution, {\it i.e.} the minimal area of the surface in $AdS_5$ ending on the WL in the (boundary) 4D Minkowski.

Now, another integrability issue enters the stage and it will be our main concern in this paper. Albeit our investigation is applicable to the full $AdS_5$ space, we shall confine our analysis to the easier $AdS_3$ subspace and its classical string solutions. In fact, the problem of finding global coordinates parametrising $AdS_3$ is equivalent to solving the two spectral linear problems $D\Psi (z,\bar z; \theta)=0$, $\bar D  \Psi (z,\bar z; \theta)=0$, with $z,\bar z$ world-sheet coordinates, $\theta$ (complex) spectral parameter and $D$, $\bar D$ linear differential operators depending on a 'potential' $\eta (z, \bar z)$ (explicitly given by formul{\ae} (\ref {D}, \ref {barD}) below). The latter must satisfy the (Lax) compatibility condition (denoting classical integrability), which in this case is a modified Sinh-Gordon equation - (\ref {cl-sinh}) below - depending uniquely on an entire function $p$ and its complex conjugate, $\bar p$. We choose $p$ to be a polynomial of degree $2N$ (integer) so that we describe light-like WLs with $4N+4$ cusps ($4N-2$ real cross ratios out of these points: the number of real parameters in $p(z)$ if the highest coefficient is $1$ and the sum of its zeroes is null) or a scattering amplitude among $4N+4$ gluons. Besides the aforementioned classical integrability, the two coupled linear problems show remarkable {\it quantum} integrability properties: in a nutshell Wronskians of solutions are $Q$ and $T$ functions (eigenvalues of some Baxter ${\bf Q}$-operator and transfer matrix respectively) of a generalised Baxter setup with moduli ({\it e.g.} the complex  coefficients of the polynomial $p(z)$). This is clearly an extension of the so-called ODE/IM correspondence \cite {ODE-IM,excited-ode-im}, which still maintains the $Q$-functions and their functional relations, the $QQ$-systems, as fundamental building blocks.

Actually, some $T$s and $Y$s are already present in previous papers \cite{YSA}, while here and in the forthcoming paper \cite {GLM-FR} we make the construction more general and extended by founding it on the basic $Q$ functions and their $QQ$ relations. Next, we derive the fundamental generalised Baxter $TQ$ relations and easily from these the Bethe Ansatz equations which are not so tractable because of their infinite number. Therefore, we convert them into a system of $2N$ non-linear integral equations (NLIEs) depending on the $2N$-ple of the coefficients of the polynomial $p(z)$. Crucially, for the whole construction of our theory we make use of two distinct discrete symmetries, $\hat \Omega $ and $\hat \Pi$, of the linear associated problems $D\Psi =0$, $\bar D \Psi =0$, which, yet, gain their effectiveness by a non-trivial action on the {\it vacuum} $\Psi$ (not invariant): in other words these are {\it broken symmetries} giving birth to the characteristic quantities of integrability ($Q$s, $T$s and $Y$s, which are generically the generators of the (local and non-local) conserved charges). Eventually, we propose an expression for the free energy in terms of solutions of the NLIEs. We claim and verify both analytically and numerically in many cases that
our expression coincides with the free energy of \cite{YSA} written in terms of solutions of Thermodynamic Bethe Ansatz (TBA) equations.

In fact, the NLIEs show many advantages with respect to TBA, at least for the polynomials we studied, which, technically speaking, are in the so-called maximal chamber \cite {YSA}: for polynomials of degree $2N$ their number is $2N$, whilst the number of TBA equations is $2N(2N-1)/2$ (quadratic grow); moreover, the procedure to write NLIEs is undoubtedly simpler than (and should shed light on) the corresponding one for TBA, which passes through $(N-1)(2N-1)$ 'wall crossings'. In addition, from the ideological point of view the NLIEs, being equivalent to the infinite Bethe Ansatz equations, lie at the heart of the integrability of the problem, and one can express in terms of their solutions all the relevant quantities, such as the $Q$ functions, the transfer matrices eigenvalues $T$ and the $Y$ functions. We will provide explicit expressions for all that at the end of Section \ref{Nlies}.

\medskip

The paper is organised as follows: in the next Section \ref{funcrel} we sketch quickly the derivation of {\it all} the functional relations which will be better analysed in \cite {GLM-FR}. Then, Section \ref{Nlies} is the core of the paper, as we derive a system of NLIEs and propose an expression for the free energy in terms of solutions of the NLIEs. Our results for the free energy are tested analytically in some simple cases (Section \ref{analcomp}) and numerically in many examples (Section \ref{numcomp}), finding perfect agreement with analogous computations made by using TBA.
Finally in Section \ref{conformal-case} we study the 'conformal limit' in which the auxiliary linear problem reduces to a Schr\"{o}dinger equation: our set of NLIEs reduces - as expected - to the NLIEs introduced in \cite{SUZ00,DDT01}.

\section{Functional relations}
\label{funcrel}

As sketched in the introduction, we start from the linear problem (the bar means complex conjugation)
\be
D\Psi =0,\,\,\, \bar D \Psi =0 \, ,
\label{ass-lin-prob}
\ee
with \footnote{We are considering $su(2)$ fundamental representations: $\sigma _i$ are Pauli matrices and $\sigma ^{\pm}=\frac{1}{2}\left (\sigma _1\pm i \sigma _2\right )$; for generalisations see for instance \cite{A22-Marian}.}
\ba
D&=&\partial _z + \frac{1}{2}\partial _z \eta \sigma ^3 -e^{\theta}\left [ \sigma ^+ e^{\eta}+\sigma ^- p(z,\vec{c}) e^{-\eta}  \right ] \, ,
\label {D}\\
\bar D&=&\partial _{\bar z} - \frac{1}{2}\partial _{\bar z} \eta \sigma ^3 -e^{-\theta}\left [ \sigma ^- e^{\eta}+\sigma ^+\bar p(z,\vec{c}) e^{-\eta} \right ] \label {barD} \, .
\ea
The polynomial
\be
p(z,\vec{c})= z^{2N} + \sum \limits _{n=0}^{2N-2}c_n z^n
\ee
depends on a $2N-1$-ple of complex coefficients $\vec{c}=(c_0,...,c_{2N-2})$ or moduli.

Zero curvature condition $[D,\bar D]=0$ constraints
the 'potential' $\eta(z,{\bar z}; \vec{c})$ to be solution of the classical modified Sinh-Gordon equation
\be
\partial _z \partial _{\bar z} \eta -e^{2\eta}+p(z,\vec{c})\bar p(z, \vec{c})e^{-2\eta}=0 \, ,
\label {cl-sinh}
\ee
To completely specify $\eta$, we impose the condition $\eta=l \ln z\bar z +O(1)$ as $|z|\rightarrow 0$, for $|l|<1/2$. Operating in such a way we are extending framework of  \cite{LZ} by introducing the moduli $\vec{c}$ (besides $c_0$) and the $AdS_3$ string problem representing the WL \cite{AM-Sinh-G-1} which is regular in $z=0$, {\it i.e.} $l=0$. Similar extension is
the series of homogeneous Sine-Gordon models \cite {HSG}.

The linear problem has two symmetries. The first one acts as
\be
\hat \Omega : \quad z \rightarrow z e^{\frac{i\pi}{N}} \, , \quad \theta \rightarrow \theta - \frac{i\pi}{N} \, , \quad \vec{c} \rightarrow \vec{c}^R \, , \quad \vec{c}^R =(c_0,c_1e^{-\frac{i\pi}{N}}, ..., c_n e^{-\frac{i\pi n}{N}}, ..., c_{2N-2}e^{\frac{2i\pi}{N}}) \, .
\label{Omega-symm}
\ee
The $\hat \Omega$ symmetry leaves invariant the polynomial $p$ and the function $\eta$:
$p(ze^{\frac{i\pi}{N}}, \vec{c}^R)=p(z, \vec{c})$, $\eta (z,\bar z;  \vec{c})=\eta (z e^{\frac{i\pi}{N}},\bar z  e^{-\frac{i\pi}{N}};  \vec{c}^R)$. However, it produces a change of sheet and then $\sqrt{p(ze^{\frac{i\pi}{N}}, \vec{c}^R)}=-\sqrt{p(z, \vec{c})}$. This symmetry involving rotations of the moduli appeared in the context of a Schr\"{o}dinger equation with polynomial potential in \cite{Suzuki1999} and, recently, in \cite{IMS,IS}: we will analyse this as limiting 2D conformal case of our massive one in Section \ref{conformal-case}.
Another symmetry of the linear operators $D$ and $\bar D$ is given by the transformation
\be
\hat \Pi : D \rightarrow \sigma ^3 D \sigma ^3 \, , \ \  \bar D \rightarrow \sigma ^3 \bar D \sigma ^3 \, ,  \quad  \theta \rightarrow \theta -i\pi \, .
\label{Pi-symm}
\ee
Now, following \cite {LZ}, we can fix two couples of solutions. The first one, $\Psi _{\pm} (z;\theta,\vec{c})$, is completely specified by the expansion around $z=\rho e^{i\varphi}=0$, $\bar z=\rho e^{-i\varphi}=0$
\be
\Psi _+ \simeq \frac{1}{\sqrt{\cos \pi l}}
\left ( \begin{array}{c} 0 \\ e^{(i\varphi +\theta )l}\end{array} \right) \, , \quad
\Psi _- \simeq \frac{1}{\sqrt{\cos \pi l}}
\left ( \begin{array}{c} e^{-(i\varphi +\theta )l} \\ 0 \end{array} \right), \,\, |z|\rightarrow 0 \,\, .
\ee
The second couple (a Jost basis) is given by the solutions $\Xi (z; \theta ,\vec{c})$ and $\Xi _1(z; \theta ,\vec{c})$ with following asymptotic leading expansion around $z=+\infty$
\be
\Xi (z; \theta ,\vec{c}) \simeq \left ( \begin{array}{c} e^{-\frac{i N \varphi}{2}} \\ -e^{\frac{iN \varphi}{2}}\end{array} \right) \exp \left [ -e^{\theta} \int ^z dz \sqrt{p(z,\vec{c})}-e^{-\theta} \int ^{\bar z} d\bar z
\sqrt{\bar p(z,\vec{c})} \right ] \, ,
\ee
\be
\Xi _1(z; \theta ,\vec{c}) \simeq -i e^{i\Phi (\theta , \vec{c})} \sigma ^3 \left ( \begin{array}{c} e^{-\frac{i N \varphi}{2}} \\ -e^{\frac{i N \varphi}{2}}\end{array} \right) \exp \left [ e^{\theta} \int ^z dz \sqrt{p(z,\vec{c})}+ e^{-\theta} \int ^{\bar z} d\bar z
\sqrt{\bar p(z,\vec{c})} \right ] \, .
\ee
We found convenient to modify the second solution by introducing the phase
\be
\label{phase}
\Phi (\theta , \vec{c})=\frac{\pi}{N} B_{-1}(\vec{c}) e^{\theta}-\frac{\pi}{N} \bar B_{-1}(\vec{c}) e^{-\theta} \, ,
\ee
where the coefficient $B_{-1}(\vec{c})$ appears in the large $|z|$ expansion of $\sqrt{p(z,\vec{c})}$:
\be
\sqrt{p(z,\vec{c})}=z^{N}+...+B_{-1}(\vec{c})z^{-1}+ O(z^{-2}) \, .
\ee
This means that the function $\Phi (\theta , \vec{c})=0$ is non zero only if $N$ is integer. Important properties of $\Phi$ are
\be
\Phi (\theta , \vec{c})=-\Phi \left ( \theta -\frac{i\pi}{N}, \vec{c}^{R}\right )=-\Phi (\theta +i\pi, \vec{c})  \label {phi-prop} \, ,
\ee
the first of which descends from $B_{-1}(\vec{c})=-e^{-\frac{i\pi}{N}}B_{-1}(\vec{c}^{R})$.

These sets of solutions transforms in a nice way under the symmetries $\hat \Omega$ and $\hat \Pi$:
\be
\hat \Omega \Psi _{\pm} (z;\theta,\vec{c})= \Psi _{\pm }\left (ze^{\frac{i\pi}{N}} ;\theta-\frac{i\pi}{N},\vec{c}^R \right ) = \Psi _{\pm} (z;\theta,c)
\ee
\be
\hat \Pi \Psi _{\pm} (z;\theta,\vec{c})=\Psi _{\pm} (z;\theta-i\pi,\vec{c})=\mp e^{\mp i\pi l} \sigma ^3
\Psi _{\pm} (z;\theta,\vec{c})
\ee
\be
\hat \Omega \Xi (z;\theta,\vec{c})= \Xi \left ( ze^{\frac{i\pi}{N}}; \theta - \frac{i\pi}{N}, \vec{c}^R \right )=\Xi _1(z;\theta,\vec{c})
\ee
\be
\left ( \hat \Omega \circ \hat \Pi \right ) \Xi (z;\theta,\vec{c})= \Xi \left ( ze^{\frac{i\pi}{N}}; \theta -i\pi - \frac{i\pi}{N}, \vec{c}^R \right )=-i \sigma ^3 e^{-i\Phi (\theta , \vec{c})}  \Xi (z;\theta,\vec{c}) \, .
\ee
Now, it is  natural to connect the two sets of solutions by means of $z$ independent coefficients $Q_{\pm}(\theta, \vec{c})$:
\be
\Xi (z;\theta, \vec{c})=Q_+(\theta ,\vec{c}) \Psi _- (z; \theta , \vec{c}) + Q_- (\theta , \vec{c}) \Psi _+ (z; \theta , \vec{c}) \label {maineq} \, .
\ee
Then, applying $\hat \Omega$ to both sides of this equation
\be
\Xi _1(z;\theta, \vec{c})=Q_+\left (\theta -\frac{i\pi}{N},\vec{c}^R \right ) \Psi _- (z; \theta , \vec{c}) + Q_- \left (\theta - \frac{i\pi}{N} , \vec{c}^R \right) \Psi _+ (z; \theta , \vec{c}),
\ee
taking the determinant of both equations and using the explicit results $\det (\Xi, \Xi _1)=-2ie^{i\Phi (\theta , \vec{c})} $ and $\det (\Psi _+,\Psi _-)=-1/\cos \pi l$, we end up with a functional relation that may be considered the foundation of the integrability of the theory, the $QQ$-system,
\be
Q_+ \left (\theta + \frac{i\pi}{2N}, \vec{c} \right ) Q_- \left (\theta - \frac{i\pi}{2N}, \vec{c}^R \right )- Q_+ \left (\theta - \frac{i\pi}{2N}, \vec{c}^R\right )Q_- \left (\theta + \frac{i\pi}{2N}, \vec{c} \right )=-2i e^{i\Phi \left(\theta + \frac{i\pi}{2N} , \vec{c}\right )} \cos \pi l  \label {Qw} \, .
\ee
This form for the $QQ$-system is very peculiar since it involves rotations of the moduli and the presence of the extra phase $\Phi$ as well.
Because of the presence of different moduli, relations (\ref {Qw}) are not closed, similarly to functional relations present at the level of $Y$-systems in \cite{MAS}.
More in general, we can think of (\ref {Qw}) as extensions of analogous relations underlying integrability in the case of ${\cal N}=4$ SYM in 4D and ${\cal N}=6$ SCS in 3D and which are often called {\it quantum spectral curves} \cite{QSC}.

Using together $\hat \Omega$ and $\hat \Pi$ on equation (\ref {maineq}) we obtain a particular form of {\it quasi-periodicity} involving the rotation of the moduli $\vec{c}$
\be
Q_{\pm } \left (\theta -i\pi - \frac{i\pi}{N} , \vec{c} ^R \right )=e^{\mp i\pi \left (l+\frac{1}{2}\right )}
e^{-i\Phi (\theta , \vec{c})}  Q_{\pm} (\theta ,  \vec{c} ) \label {qper} \, .
\ee
Quasi-periodicity (\ref {qper}) allows to bring the $QQ$-system (\ref {Qw}) in a form which does not involve rotations of the moduli and the phase $\Phi$:
\be
e^{i\pi l} Q_+(\theta ,\vec{c})Q_-(\theta +i\pi, \vec{c}) + e^{-i\pi l}Q_-(\theta ,\vec{c} )Q_+(\theta +i\pi, \vec{c})=-2\cos \pi l
\, .
\ee
Another useful property of $Q_{\pm }$ is its behaviour under complex conjugation
\be
Q_{\pm}(\theta , \vec{c} )=-\bar Q_{\mp} (-\bar \theta , \vec{c} ) \label {qcompl} \, ,
\ee
which derives from the relations
\be
\sigma ^1 \bar \Xi (z;-\theta , \vec{c})=-\Xi (z;\bar \theta , \vec{c}) \, , \quad
\sigma ^1 \bar \Psi _{\pm}(z;-\theta , \vec{c})=\Psi _{\mp}(z;\bar \theta , \vec{c}) \, .
\ee
In addition, usual real-analyticity
\be
\bar Q_{\pm}(\theta , \vec{c})=Q_{\pm}(\bar \theta , \bar \vec{c}) \, , \label {re-an}
\ee
holds.

In order to write NLIEs we need to introduce the transfer matrix $T$, which we define as a quadratic construct of $Q$'s:
\be
T (\theta , \vec{c})=
\frac{i}{2\cos \pi l} \left [ e^{-2i\pi l} Q_+(\theta +i\pi, \vec{c})Q_-(\theta -i\pi, \vec{c})-e^{2i\pi l} Q_+(\theta -i\pi, \vec{c})Q_-(\theta +i\pi, \vec{c}) \right] \, .
\label{trmat}
\ee
Using the $QQ$ system, we get the $TQ$-relation or $TQ$-system,
\be
T (\theta , \vec{c}) Q_{\pm}(\theta ;\vec{c})= e^{\pm i\pi \left (l+\frac{1}{2}\right)}  Q_{\pm} \left (\theta -i\pi ; \vec{c} \right) + e^{\mp i\pi \left (l+\frac{1}{2}\right)} Q _{\pm}\left (\theta + i\pi ; \vec{c} \right ) \, .
\label{closed-TQ}
\ee
Importantly, it follows from their construction that both $T (\theta , \vec{c})$ and  $Q_{\pm}(\theta ;\vec{c})$ are analytic functions of $\theta$. This information will be used in next Section
to derive in quite a standard way a system of Bethe equations and NLIEs.

\medskip

We stress that that the construction of $QQ$ and $TQ$-systems was possible thanks to the use of the $\hat \Omega $ symmetry. This symmetry was not used in \cite{YSA}, but we underline its importance to unveil the
basics of integrability of the problem. More importantly, $\hat \Omega$ symmetry is essential when differential operators have two irregular singularities, like in Liouville ODE \cite{FG-pure-matt}.

\medskip

The last ingredient for the derivation of NLIEs is the leading behaviour of $Q_{\pm}$ when $\textrm{Re}\theta \rightarrow \pm \infty$. This is obtained by studying the linear problem (\ref{ass-lin-prob}) at large $\theta$. Details will be reported in the publication \cite {GLM-FR}. We give here only the final result:
\be
\ln Q_{\pm} \left ( \theta +i\pi \frac{N+1}{2N};\vec{c}\right )
 \sim  -(w_0(\vec{c})+\alpha (\vec{c}) )e^{\theta}-(\bar w_0(\vec{c})+\bar \alpha (\vec{c}))e^{-\theta}
\label {largetheta} \, ,
\ee
with
\be
w_0(\vec{c})=-\int _{0}^{+\infty} dx \left [ \sqrt{P(x,\vec{c})}-q_N(x,\vec{c}) \right ] \, , \quad
\alpha (\vec{c})=\frac{\pi}{2N} B_{-1}(\vec{c})e^{\frac{i\pi}{2N}} \, .
\label {apex1}
\ee
The function $P(x,\vec{c})$ is a polynomial, related to $p(x,\vec{c})=\sum _{n=0}^{2N} c_n x^n$ by a change of sign and a half rotation of coefficients $\vec{c}$:
\be
P(x,\vec{c})=\sum _{n=0}^{2N} c_n e^{i\pi\frac{2N-n}{2N}} x^n \, .
\ee
The function $q_N$ is another polynomial, defined by the condition
\be
\sqrt{P(x,\vec{c})}=q_N(x)+o(1/x) \, , \quad x \rightarrow +\infty \, ,
\ee
which assures convergence of the integration in (\ref{apex1}).

\section{NLIEs}
\label{Nlies}
\setcounter{equation}{0}

By using the $TQ$ system (\ref {closed-TQ}), the quasiperiodicity property of $Q$ (\ref {qper}) and its asymptotic behaviour (\ref {largetheta}), we can write a set of NLIEs.
As a preliminary step, we introduce the function $Q(\theta ; k, \vec{c})$ which, when $\textrm{sgn} k=\pm$, equals $Q_{\pm} (\theta ; \vec{c})$, with $l=\pm 2k -\frac{1}{2}$, respectively. Then, we may think of extending (3.24) of \cite {LZ} by writing $Q(\theta ; k, \vec{c})$ as a product over its zeroes which implements \footnote {The next formula (\ref {Qprop}) is not the only possible with these implementations, but it seems to us the simplest one.} quasiperiodicity
and property (\ref {qcompl}), i.e. $Q(\theta ; k,\vec{c})=-\bar Q(-\bar \theta ; -k,\vec{c})$:
\ba
Q(\theta ; k, \vec{c})&=&e^{\frac{\theta N}{\pi(N+1)}\Phi (\theta , \vec{c})}C(k) e^{\frac{2\theta N k}{1+N}}
\prod _{p=0}^{2N-1} \prod _{n=0}^{+\infty} \left ( 1-e^{\frac{\theta }{1+N}}e^{-\frac{i\pi p}{N}}e^{-\frac{\theta _n(\vec{c}^{R^{p}};k)}{1+N}} \right )\, \cdot \nonumber \\
&\cdot & \prod _{p=0}^{2N-1}\left ( 1-e^{-\frac{\theta }{1+N}}e^{\frac{i\pi p}{N}}e^{-\frac{\bar \theta _n(\vec{c}^{R^{p}};-k)}{1+N}} \right )
 = e^{-i\Phi (\theta , \vec{c})}Q \left (\theta + i\pi +\frac{i\pi}{N}; k, \vec{c}^{R^{-1}}\right ) e^{-2i\pi k} \, ,
\label {Qprop}
\ea
with $C(k)=-C(-k)$ a real constant and
$\textrm{Re}\theta _n(\vec{c}^{R^{p}};k)$ positive. In the case of \cite {LZ}, i.e. $c_n=0, n=1,...,2N-2$, $\theta _n(\vec{c}^{R^{p}};k)$ are real and do not depend on $p$: then, formula (\ref {Qprop}) reduces to (3.24) of \cite {LZ}.
Remembering the form of the $p$ times rotated $2N-1$-ple
\be
\vec{c}^{R^{p}}=(c_0,c_1e^{-\frac{i\pi p}{N}},..., c_n e^{-\frac{i\pi p n}{N}}, ..., c_{2N-2} e^{\frac{2i\pi p}{N}})\, ,
\ee
we expect $2N$ NLIEs, since $2N$ is the number of discrete rotations (including the identity, i.e. the number of independent sets of $\vec{c}^{R^{p}}$) of the group acting on the moduli space. In order to write the NLIEs, we recall the $TQ$-system (\ref {closed-TQ})
\be
T (\theta ,  \vec{c}) Q (\theta ; k,  \vec{c})= e^{2i\pi k}  Q \left (\theta -i\pi ; k,\vec{c} \right) + e^{-2i\pi k} Q \left (\theta + i\pi ; k,  \vec{c} \right ) \, ,
\label{closed-TQk}
\ee
which indicates, as usual, that the zeroes of $Q$ are indeed the Bethe roots, $\theta _n(\vec{c};k)$. This suggests us the definition of their counting function \cite{DDV}
\be
Z(\theta ; k,  \vec{c})= i \ln \left [ \frac{Q(\theta -i\pi;k, \vec{c})}{Q(\theta +i\pi;k, \vec{c})} e^{4\pi i k} \right ] \, , \label {count}
\ee
so that the Bethe Ansatz equations take the form
\be
e^{iZ\left (\theta _n(\vec{c}^{R^{p}};k);k, \vec{c}^{R^{p}}\right )}=-1 \, , \,
e^{iZ\left (-\bar \theta _n(\vec{c}^{R^{p}};-k);k, \vec{c}^{R^{p}}\right )}=-1 \, , \,p=0,...,2N-1 \, ,
\label{BAEqs}
\ee
and equivalent ones obtained by use of the periodicity of the counting function $Z(\theta ; k,  \vec{c})=Z\left (\theta + i\pi +\frac{i\pi}{N}; k, \vec{c}^{R^{-1}}\right )$.
Therefore it is very relevant to recast the counting function in a more explicit form as
\be
Z(\theta ; k, \vec{c})= - \frac{4\pi k+2N \Phi (\theta , \vec{c}) }{1+N}+i\ln \left (  \prod _{m=0}^{2N-1} \prod _n \frac{\sinh \left [ \frac{i\pi-\theta + \theta _n(\vec{c}^{R^m};k) }{2(1+N)} +\frac{i\pi m}{2N} \right ] \sinh \left [ \frac{i\pi-\theta -\bar \theta _n(\vec{c}^{R^m};-k)}{2(1+N)} +\frac{i\pi m}{2N} \right ]}{\sinh \left [ \frac{\theta +i\pi-\theta _n(\vec{c}^{R^m};k)}{2(1+N)} - \frac{i\pi m}{2N} \right ] \sinh \left [ \frac{\theta +i\pi+\bar \theta _n(\vec{c}^{R^m};-k)}{2(1+N)} - \frac{i\pi m}{2N} \right ] } \right ) \, .
\ee
In order to understand the position of the Bethe roots we refer to the behaviour for large $\theta $ of $Z(\theta ; k,  \vec{c}^{R^m})$. We claim that, when $\theta \rightarrow \pm \infty$,
\ba
&& Z(\theta ; k,  \vec{c}^{R^m}) \simeq \frac{r(\vec{c}^{R^m})}{2}e^{\theta}-\frac{\bar r(\vec{c}^{R^m})}{2}  e^{-\theta} = |r(\vec{c}^{R^m})| \sinh (\theta + i \phi _m) \, ,
\label {Zlargetheta}  \\
&& r(\vec{c}^{R^m})=-2w_0(\vec{c}^{R^m})e^{-\frac{i\pi}{2N}} -2 w_0(\vec{c}^{R^{m-1}})e^{\frac{i\pi}{2N}}+\delta r (\vec{c}^{R^m}) =  r_0(\vec{c}^{R^m})
+\delta r (\vec{c}^{R^m})\, , \label {rdefinitions} \\
&& \delta r(\vec{c})=-\frac{2\pi}{N} B_{-1}(\vec{c}) \, ,
\ea
where we introduced
the adimensional quantity $r(\vec{c}^{R^m})=|r(\vec{c}^{R^m})| e^{i\phi _m}$ and used the property $\bar w_0(\vec{c})=w_0(\bar \vec{c}^{R^{-1}})$.
Relation (\ref {Zlargetheta}) comes from (\ref {count}, \ref {largetheta}).

From these definitions it follows that the quantity $r(\vec{c})$ is obtained by an integration
on a contour $\Gamma $ defined by $]-i\infty , -i0 ] \bigcup [0, +\infty [ $:
\be
r(\vec{c}^{R})=2 e^{\frac{i\pi}{2N}} \int _{\Gamma} dz \left [ \sqrt{P(z,\vec{c})}- q_N(z,\vec{c})\right ]
\label{rwedge} \, .
\ee
In addition, from the asymptotic behaviour (\ref {Zlargetheta}) we deduce that, for very large $\theta _n$, $\textrm{Im}\, \theta _n(\vec{c}^{R^m};k)=
-\textrm{Im}\, \bar \theta _n(\vec{c}^{R^m};-k)=-\phi _m$.
We suppose that for finite $\theta _n$ the position of the Bethe roots does not displace much from such lines.

Then, a general sum over the Bethe roots $\theta _n(\vec{c}^{R^m};k)$, $-\bar \theta _n(\vec{c}^{R^m};-k)$ can be converted into integrals by using Cauchy theorem \cite {DDV}, which fact for $Z$ itself boils down, after standard manipulations, to
\ba
Z(\theta ;k, \vec{c})&=&   - \frac{4\pi k+2N \Phi (\theta , \vec{c}) }{1+N}+\sum _{m=0}^{2N-1} \int _{\textrm{Im}x=-\phi _m} dx K_m(\theta -x) Z(x ; k, \vec{c}^{R^m})- \nonumber \\
&-& 2 \sum _{m=0}^{2N-1} \int _{\textrm{Im}x=-\phi _m} dx \left [ K_m(\theta -x -i\epsilon )L_+(x ; k,  \vec{c}^{R^m})- K_m(\theta -x +i\epsilon )L_-(x ; k,  \vec{c}^{R^m})\right ] \, , \nonumber
\ea
with $\epsilon >0$ and $L_{\pm}(x;k, \vec{c})=\frac{1}{2i}\ln \left [1+e^{\pm iZ(x\pm i\epsilon ;k, \vec{c})}\right ]$. The positive number $\epsilon $ is chosen in such a way that integration contour contains all the Bethe roots, i.e.
$|\textrm{Im}\, \theta _n(\vec{c}^{R^m};k)+\phi _m|<\epsilon$, $\forall n,m$.

The kernels $K_m(x)$ equal
\be
K_m(x)=\frac{1}{2\pi i} \frac{d}{dx} \ln \left (
\frac{\sinh \left [ \frac{i\pi +x }{2(1+N)} -\frac{i\pi m}{2N} \right ] }{\sinh \left [ \frac{i\pi-x}{2(1+N)} +\frac{i\pi m}{2N} \right ] } \right )=K_{m+2N}(x) \, ,
\ee
and in Fourier transform read
\be
\hat K_0(p)=-\frac{\sinh N\pi p}{\sinh (1+N)\pi p} \, , \quad
\hat K_m(p)= \frac{\sinh \pi p}{\sinh (1+N)\pi p}e^{(1+N)\pi p \left (\frac{m}{N}-1\right )}
\quad \textrm{for} \quad 1\leq m \leq 2N-1 \, .
\ee
Using the periodicity $K_m(x)=K_{m+2N}(x)$ of the kernels and bringing the integrations on the real axis, we can also write
\ba
&& Z(\theta -i\phi _n ;k, \vec{c}^{R^n})= - \frac{4\pi k+2N \Phi (\theta-i\phi _n, \vec{c}^{R^n}) }{1+N}+\sum _{m=0}^{2N-1} \int _{-\infty}^{+\infty} dx K_{m-n}(\theta -x+i\phi_m-i\phi_n) Z(x-i\phi_m ; k, \vec{c}^{R^m})- \nonumber \\
&& -2 \sum _{m=0}^{2N-1} \int _{-\infty}^{+\infty} dx \Bigl [ K_{m-n}(\theta -x +i\phi_m-i\phi_n-i\epsilon )L_+(x-i\phi_m ; k,  \vec{c}^{R^m})- \nonumber \\
&&- K_{m-n}(\theta -x+i\phi_m-i\phi_n +i\epsilon )L_-(x-i\phi_m ; k,  \vec{c}^{R^m})\Bigr ] \, .  \nonumber
\ea
Passing to Fourier transforms, we define the matrix $\hat M_{i,j}(p)=\delta _{i,j}-e^{p\phi_i}\hat K _{j-i}(p) e^{-p\phi _j}$, whose determinant is explicitly computed:
\be
\textrm{det} \hat M (p) =\prod _{m=0}^{2N-1} \left (1-\sum _{n=0}^{2N-1} \hat K_n (p) e^{\frac{i\pi nm}{N}} \right ) = \left ( 2 \cosh \frac{\pi p}{2} \right )^{2N} \frac{\sinh \pi p}{\sinh \pi p (1+N)} \, .
\ee
The inversion of $\hat M_{i,j}(p)$ allows to write the final set of NLIEs. The driving term for the $n$-th equation
equals
\be
-2\pi k - \Phi (\theta -i\phi _n,  \vec{c}^{R^n})=-2\pi k - \frac{\pi}{N} \left ( B_{-1}(\vec{c}) e^{i\pi\frac{N+1}{N}n} e^{\theta -i\phi _n}-\bar B_{-1}(\vec{c}) e^{i\pi\frac{N-1}{N}n} e^{-\theta +i\phi _n} \right ) \label{driv1}
\ee
plus zero modes of the matrix $\hat M_{i,j}$, i.e. solutions of the homogeneous equation $\hat M_{i,j}c_j=0$.
In this respect, we remark the following property
\be
 \left (1-\sum _{n=0}^{2N-1} \hat K_n (p) e^{\frac{i\pi nm}{N}} \right )  \delta (p\pm i)=0 \, ,
\ee
which holds for all $m=0,...,2N-1$, except when $m=N-1$ for $\delta (p-i)$ and for $m=N+1$ for $\delta (p+i)$ (therefore, this two 'problems' appear only for integer $N$). This means that zero modes are linear combinations of
vectors with the form $c^{(\pm,m)}_n= e^{\mp i \phi _n}e^{\frac{i\pi nm}{N}} \delta (p\pm i)$, with $m=0,...,2N-1$ labeling the independent solutions - with the two exceptions mentioned before. We remark that the quantities $r_0(\vec{c}^{R^n})e^{\theta -i\phi _n}$,
$\bar r_0(\vec{c}^{R^n})e^{-\theta +i\phi _n}$ can be constructed out of the zero modes, i.e. we have the relation
\be
\sum_{n=0}^{2N-1}e^{-\frac{i\pi (N+1)}{N}n} r_0(\vec{c}^{R^n})=0 \, ,
\ee
which we have numerically verified, see Table \ref{r0-z4-z2-z-1} and Table \ref{quartic-z-lg-r0} in Section \ref{numcomp}.
Adding then the last two terms of (\ref {driv1}) coming from the extra phase $\Phi$, we completely reconstruct the asymptotic behaviour (\ref {Zlargetheta}) previously announced, i.e.
\be
Z(\theta -i\phi _n; k,  \vec{c}^{R^n})\sim | r(\vec{c}^{R^n})|\sinh \theta
\label {Zlargethetashift} \, .
\ee
Putting everything together, the final equations are ($0\leq n \leq 2N-1$):
\ba
Z(\theta -i\phi_n ; k,  \vec{c}^{R^n})&=&  | r(\vec{c}^{R^n})|\sinh \theta -2\pi k +
2 \sum _{m=0}^{2N-1} \int _{-\infty}^{+\infty} dx \Bigl [ G_{m-n}^{(N)}(\theta -x-i\phi _n+i\phi _m-i\epsilon ) L_+(x-i\phi _m ; k,  \vec{c}^{R^m}) - \nonumber
 \\
&-&  G_{m-n}^{(N)}(\theta -x-i\phi _n+i\phi _m+i\epsilon ) L_-(x-i\phi _m ; k,  \vec{c}^{R^m}) \Bigr ]
\label {nlies}
\, .
\ea
We now give the explicit form for the kernels: the periodicity $G^{(N)}_m(x)=G^{(N)}_{m+2N}(x)$ is understood. For small $N$ we have ($1\leq m \leq 2N-1$):
\be
\hat G^{(1/2)}_0(p)=\frac{1}{4\cosh ^2 \frac{\pi p}{2}} \, , \quad \hat G^{(1)}_0(p)=-\hat G^{(1)}_1(p)=\frac{1}{4\cosh ^2 \frac{\pi p}{2}} \nonumber
\ee
\be
\hat G^{(3/2)}_0(p)= \frac{1}{\left (2\cosh \frac{\pi p}{2}\right )^2} \, , \quad \hat G^{(3/2)}_m(p)=- \frac{1}{\left (2\cosh \frac{\pi p}{2}\right )^2} e^{\pi p \left (\frac{2m}{3}-1\right )} \, . \nonumber
\ee
We can then guess ($1\leq m \leq 2N-1$) the general formul{\ae}:
\ba
\hat G^{(N)}_0(p)&=& \frac{1}{\left (2\cosh \frac{\pi p}{2}\right )^2} \quad \Rightarrow \quad
G^{(N)}_0(x)=\frac{1}{2\pi^2} \frac{x}{\sinh x} \nonumber \\
\hat G^{(N)}_m(p)&=& -\frac{1}{\left (2\cosh \frac{\pi p}{2}\right )^2} e^{\pi p \left (\frac{m}{N}-1\right )}\quad \Rightarrow \quad G^{(N)}_m(x)=\frac{1}{2\pi^2} \frac{x-i\pi\left (\frac{m}{N}-1\right)}{\sinh \left (x-i\pi \frac{m}{N}\right )} \, . \label{kerG}
\ea
Equations (\ref {nlies}) with kernels (\ref {kerG}) are valid in the strip defined by the conditions $|\textrm{Im} \theta -\phi _n +\phi _m \pm \epsilon| < \textrm{min} \{ \pi , \pi/N \}$, $\forall n,m$. They can be solved analytically in few cases, which we report in next Section, or alternatively by numerical iteration: we will discuss many examples in Section \ref{numcomp}.
We observe the simple summation
\be
\sum _{m=0}^{2N-1} \hat G_m(p)=\frac{\sinh \left ( \frac{\pi p}{2N}-\frac{\pi p}{2}\right )}{2\cosh \frac{\pi p}{2}\sinh \frac{\pi p}{2N}}=\hat G^{LZ}(p)|_{\alpha =N} \, ,
\ee
where $\hat G^{LZ}(p)|_{\alpha =N}$ is the Fourier transform of the kernel of equation (3.32) of \cite{LZ} (and of (2.29) of \cite {FR03}) to which our equations (\ref {nlies}) reduce in the case in which $Z(\theta ; k,  \vec{c}^{R^n})$ do not depend on $n$.

Once we know $Z$ we can very reasonably conjecture that the free energy be given by the functional \footnote {Interestingly, the property $\bar Z(\theta; k, \vec{c})=-Z(-\bar \theta ; -k , \vec{c})$, which descends from  (\ref {qcompl}), is enough to ensure that the quantity given by (\ref {Free}) is real.}
\be
F=\sum _{n=0}^{2N-1} \int _{-\infty}^{+\infty }\frac{d\theta}{2\pi}[ f_n(\theta +i\epsilon ) L_+(\theta-i\phi _n;k,\vec{c}^{R^n}) -  f_n(\theta -i\epsilon ) L_-(\theta-i\phi _n;k,\vec{c}^{R^n})] + (k\rightarrow -k)  \, ,
\label {Free}
\ee
with $f_n(\theta)= |r(\vec{c}^{R^n})| \sinh \theta$, so that when
$l=0$ it does coincide with the TBA contribution \cite {YSA,HISS}
\be
A_{free}= \sum _{s=1}^{N_{max}} \int^{+\infty}_{-\infty} \frac{d\theta}{2\pi} |m_s| \cosh \theta \ln \left [ 1+ \tilde Y_s^{AMSV}(\theta) \right ] \, , \label{Afree-tba}
\ee
to the area (scattering amplitude) $A=A_{free}+A_{periods}$ of $4N+4$ gluons. The functions $\tilde Y_s^{AMSV}(\theta)$, whose number $N_{max}$ depends on the parameters $\vec{c}$ and varies from $2N-1$ (minimal chamber) to $(2N-1)N$ (maximal chamber), satisfy TBA equations. For parameters in minimal chamber TBA equations have the form (26) of \cite {YSA}:
\ba
&& \ln \tilde Y_s^{AMSV}(\theta )=-|m_s|\cosh \theta + \sum _{s'=1}^{2N-1}\int d\theta ' K_{s,s'}(\theta -\theta ')\ln (1+ \tilde Y_{s'}^{AMSV}) (\theta ') \, , \label{TBAeqs}\\
&& \quad K_{s,s'}(\theta )=\frac{\delta _{s,s'+1}+\delta_{s,s'-1}}{2\pi \cosh (\theta +i\varphi _s^{TBA} -i\varphi _{s'}^{TBA})} \, , \quad s=1,...,2N-1 \, . \nonumber
\ea
In next Sections, we test the conjectured coincidence in examples where analytic or numerical computations can be done.
For what concerns the precise connexion with TBA equations of \cite {YSA,HISS}, this will be pursued in more details in a future publication.
For the moment we anticipate the relation between the functions $Y_{1}^{AMSV}(\theta)=\tilde Y_{1}^{AMSV}(\theta -i\varphi_{1})$ of \cite {YSA} and the counting function:
\be
1+Y_{1}^{AMSV}(\theta)=\left [1+e^{-iZ\left(\theta - \frac{i\pi}{2}; \pm \frac{1}{4},\vec{c}\right)}\right] \left [1+e^{iZ\left(\theta + \frac{i\pi}{2}; \pm \frac{1}{4},\vec{c}\right)}\right] \, . \label {Y-Znew}
\ee
Before ending this part, we want to stress that NLIEs (\ref {nlies}) are valid if $|c_n|$, $n=1,...,2N-2$ and, consequently the phases $\phi _n$, are not too large: to be precise, remembering what we wrote just after equation (\ref {kerG}), the condition $|\phi _n-\phi _m| <\textrm{min}\{ \pi , \pi/N \}$, $\forall n,m$  has to be satisfied. In this case, we have $2N$ NLIEs in the form (\ref {nlies}). For general $c_n$ the form of the NLIEs changes and we expect the occurrence of a phenomenon analogous to the 'wall crossing' of TBA equations \cite {YSA,GMN,HISS}: in the context of integrability this phenomenon already appeared and was called 'second determination' of the NLIEs \cite{DDV}.

\medskip

As we discusses in the introduction, the main importance of the NLIEs in the context of integrability is that one can express in terms of the counting functions all the relevant functions of the problem. The connection to the $Y$-functions descends from (\ref  {Y-Znew}).  
On the other hand, the transfer matrix is a functional of the counting functions. For instance,
in the simplest case, in which there is no dependence on the moduli, for the ground state eigenvalue we find, following procedures discussed in \cite{FR03a,FR03},
\be
T(\theta)=e^{\frac{2i\pi k}{1+N}}\exp [F(\theta ;k)]+ e^{-\frac{2i\pi k}{1+N}}\exp \left [-F\left (\theta -i\frac{\pi}{N} ;k \right) \right ] \, ,
\ee
where, when $0<\textrm{Im}\theta <\pi$,
\be
F(\theta ;k)=r \tan \frac{\pi}{2N} \cosh \theta - \int _{-\infty}^{+\infty} \frac{d\theta '}{2i\pi} \frac{1}{\sinh (\theta -\theta')} \ln \frac{1+e^{iZ(\theta '+i\epsilon;k)}}{1+e^{-iZ(\theta '-i\epsilon;k)}} \, ,
\ee
and, when $-\pi/N < \textrm{Im}\theta <0$,
\be
F(\theta ;k)=-i \frac{r}{\cos \frac{\pi}{2N}}\cosh \left (\theta +\frac{i\pi}{2N}\right ) +2\pi i k +
\int  _{-\infty}^{+\infty} \frac{d\theta '}{2i\pi} G_{T}(\theta -\theta ') \ln \frac{1+e^{iZ(\theta '+i\epsilon;k)}}{1+e^{-iZ(\theta '-i\epsilon;k)}} \, ,
\ee
with
\be
G_T(x)=\int  _{-\infty}^{+\infty} dk e^{-ikx+\frac{\pi k}{2N}} \frac{i \tanh \frac{\pi k}{2}}{\sinh \left (\frac{\pi k}{2}+\frac{\pi k}{N} \right )} \, .
\ee
Finally, the ground state eigenvalue of the $Q$ function in the simplest case with no moduli has been written in formul{\ae} (3.38-3.40) of \cite{LZ} and therefore we refer to that paper for explicit expressions.

Formul{\ae} for the eigenvalues of the $T$ and $Q$ functions in terms of $Z$ in the general case with arbitrary moduli are only technical complications and therefore for the moment we do not write them.


\section{Analytic computations}
\setcounter{equation}{0}
\label{analcomp}

{\bf Regular polygons}

The regular polygon case corresponds to polynomials
\be
p(z)=z^{2N} \, .
\ee
This means that $c_{2N}=1$, whilst all the other moduli $c_n=0$, $n=0,1,...,2N-1$.
This is the 'massless' limit $w_0(\vec{c})=0$: in this limit, the evaluation of the free energy is
done by using the 'dilogarithmic trick' applied to NLIEs (see \cite {DDV}).
We have that
\be
F=2N \frac{\pi}{6}c_{eff}  \label {free-en} \, ,
\ee
where the effective central charge $c_{eff}$
is easily computed (following Section 7.1 of the first of \cite {DDV}) as
\be
c_{eff}=1-\frac{3}{2(1+N)}=\frac{2N-1}{2N+2} \, \label {c-eff} \, .
\ee
Then, from (\ref {c-eff}) and (\ref {free-en}) we get the free energy
\be
F=\frac{\pi}{6}2N \frac{2N-1}{2N+2} \label {free} \, ,
\ee
which agrees with (4.7) of \cite {HISS}, since the number $n$ there is expressed by our $N$ by
\be
n=2N+2 \, .
\ee

{\bf The octagon}

Another case in which explicit analytic computation is possible is the octagon: $N=1$, $p(z)=z^2+c_0$.
This case corresponds in the ODE limit to the harmonic oscillator and the solutions of the two NLIEs are\footnote{In this case the quantity $r_0$ which is constructed from the zero modes vanishes and the driving term is due only to the additional term $\delta r$.}
\be
Z(\theta ; k,\vec{c})= Z(\theta; k, \vec{c}^R)=\pi |c_0|\sinh \theta -2\pi k \, , \quad
\vec{c}=(c_0,0,1) \, .
\ee
In order to evaluate the free energy it is useful to rewrite (\ref {Free}) as:
\be
F=\pi |c_0| \textrm{Im} \int \frac{d\theta}{2\pi}\sinh (\theta +i\epsilon) 2 \ln \left [1+e^{i\pi |c_0| \sinh (\theta +i\epsilon)-2i\pi k} \right ] +(k\rightarrow -k) \, .
\ee
We stick to the case $l=0$, that is $k=1/4$. Then,
we choose the arbitrary constant $\epsilon =\pi/2$: with these positions
\be
F=\pi |c_0| \textrm{Im} \int \frac{d\theta}{2\pi} i \cosh \theta \ 2 \ln \left [1+e^{-2\pi |c_0|\cosh \theta} \right ] \, .
\ee
Expanding the logarithm in series, we find the result as a series of Bessel functions $K_1$:
\be
F=
-2|c_0| \sum  _{n=1}^{+\infty} \frac{(-1)^n}{n} K_1(2\pi |c_0| n) \, ,
\ee
which agrees with computations made \cite {AMapr09} using minimal area.

\section{Numerical computations}
\setcounter{equation}{0}
\label{numcomp}

We report some numerical computations of free energy.
We test our NLIEs against the TBA equations in \cite{YSA, IMS}. Let us first consider the polynomial $p(z,\vec{c})=z^4+c_2 z^2+c_1z-1$, where $c_1$ and $c_2$ are chosen to keep the TBA equations in maximal chamber. In this case, the TBA equations are given by\footnote{In our notation, the masses $m_a$ are twice of the masses in \cite {IMS}.}\be
\begin{aligned}
\label{6tba}
\widetilde{\epsilon}_{1}&=|m_{1}|\cosh\theta-K_{1,2}\star\widetilde{L}_{2}-K_{1,12}^{+}\star\widetilde{L}_{12}-K_{1,23}^{+}\star\widetilde{L}_{23}-K_{1,123}^{+}\star\widetilde{L}_{123},\\\widetilde{\epsilon}_{2}&=|m_{2}|\cosh\theta-K_{2,1}\star\widetilde{L}_{1}-K_{2,3}\star\widetilde{L}_{3}-K_{2,12}\star\widetilde{L}_{12}-K_{2,23}\star\widetilde{L}_{23}-2K_{2,123}\star\widetilde{L}_{123},\\\widetilde{\epsilon}_{3}&=|m_{3}|\cosh\theta-K_{3,2}\star\widetilde{L}_{2}-K_{3,12}^{+}\star\widetilde{L}_{12}-K_{3,23}^{+}\star\widetilde{L}_{23}-K_{3,123}^{+}\star\widetilde{L}_{123},\\\widetilde{\epsilon}_{12}&=|m_{12}|\cosh\theta-K_{12,1}^{-}\star\widetilde{L}_{1}-K_{12,3}^{-}\star\widetilde{L}_{3}-K_{12,2}\star\widetilde{L}_{2}-K_{12,123}^{-}\star\widetilde{L}_{123},\\\widetilde{\epsilon}_{23}&=|m_{23}|\cosh\theta-K_{23,1}^{-}\star\widetilde{L}_{1}-K_{23,3}^{-}\star\widetilde{L}_{3}-K_{23,2}\star\widetilde{L}_{2}-K_{23,123}^{-}\star\widetilde{L}_{123},\\\widetilde{\epsilon}_{123}&=|m_{123}|\cosh\theta-K_{123,1}^{-}\star\widetilde{L}_{1}-K_{123,3}^{-}\star\widetilde{L}_{3}-2K_{123,2}\star\widetilde{L}_{2}-K_{123,12}^{+}\star\widetilde{L}_{12}-K_{123,23}^{+}\star\widetilde{L}_{23}.
\end{aligned}
\ee
where $\tilde \epsilon _j=-\ln \tilde Y_j^{AMSV}$, $\tilde L_j=\ln \left (1+\tilde e^{-\epsilon _j} \right )$ and $\star$ denotes the convolution. The kernels in the integral equations are
\be
K_{r,s}=\frac{1}{2\pi}\frac{1}{\cosh\big(\theta+i({\rm Arg}(m_s)-{\rm Arg}(m_r))\big)}.
\ee
The superscripts of kernel denote the shift of the argument $f^{\pm}(\theta)=f(\theta\pm \frac{\pi i}{2})$. $A_{\rm free}$ is now given by
\be
\begin{aligned}
A_{{\rm free}}&=\frac{1}{2\pi}\int_{-\infty}^{\infty}\Big(|m_{1}|\cosh\theta\log\big(1+e^{-\tilde{\epsilon}_{1}}\big)+|m_{2}|\log\big(1+e^{-\tilde{\epsilon}_{2}}\big)+|m_{3}|\cosh\theta\log\big(1+e^{-\tilde{\epsilon}_{3}}\big)\\&\quad+|m_{12}|\log\big(1+e^{-\tilde{\epsilon}_{12}}\big)+|m_{23}|\log\big(1+e^{-\tilde{\epsilon}_{23}}\big)+|m_{123}|\log\big(1+e^{-\tilde{\epsilon}_{123}}\big)\Big)d\theta.
\end{aligned}
\ee
When $c_1=0$, we get the double well potential, where $m_1=m_3$ and $m_{12}=m_{23}$.
The $r$ in NLIEs can be computed explicitly by using (\ref{rdefinitions}). In Table \ref{r-quartic-near-LZ}, we list several examples for $r(\vec{c})$ and $r(\vec{c}^{R_1})$. Other $r(\vec{c})$ are obtained by $r(\vec{c}^{R_2})=r(\vec{c})$ and $r(\vec{c}^{R_3})=r(\vec{c}^R)$.  The free energy can be computed by solving the NLIEs (\ref{nlies}) and then by substituting the numerical solution into equation (\ref{Free}) for the free energy. In Table \ref{quartic-max} we show the numerical results of $\frac{6}{\pi}F$ and $\frac{6}{\pi}A_{\rm free}$ for several values of $c_2$. As expected, $F$ matches $A_{\rm free}$ with a very high precision.
\begin{table}[H]
\caption{\footnotesize   $r(\vec{c})$ and $r(\vec{c}^{R_1})$: $p(z,\vec{c})=z^4+c_2 z^2-1$. Other $r(\vec{c})$ are obtained by $r(\vec{c}^{R_2})=r(c)$ and $r(\vec{c}^{R_3})=r(\vec{c}^R)$. }
\begin{center}
\begin{tabular}{c|c|c}
$c_2$&$r(\vec{c})$ & $r(\vec{c}^{R_1})$\\\hline
$1$ & $2.582636248747118`  $ & $5.049379803768095` $\\\hline
$\frac{1}{2}$ & $2.973818997047456` $ & $4.181170379881933` $ \\\hline
$0$ &$3.4960767390561593` $ &$3.4960767390561593` $ \\\hline
\end{tabular}
\end{center}
\label{r-quartic-near-LZ}
\end{table}%
\begin{table}[H]
\caption{\footnotesize  Numeric check in maximal chamber: $p(z,\vec{c})=z^4+c_2 z^2-1$. The NLIEs are solved by Fourier discretisation with cutoff $(-12,12)$ and $2^{12}$ points. The parameter $\epsilon$ in eq.(3.15) is fixed to be $1/2^4$. The TBA equations are solved by $2^{14}$ points and cutoff $(-20,20)$.}
\begin{center}
\begin{tabular}{c|c|c|c}
$c_2$&$\frac{6}{\pi}F$ & $\frac{6}{\pi}A_{\rm free}$& $F/A_{\rm free}$\\\hline
$1$ & $0.03710467901895455`$ & $0.03710480745177178`$&$0.9999965386475216`$ \\\hline
$\frac{1}{2}$ & $0.04910296244435445`$ & $0.049102990179460966`$&$0.9999994351646119`$ \\\hline
$0$ &$0.055363520349671534` $ &$0.0553655493634881` $&$0.9999633524124677`$ \\\hline
\end{tabular}
\end{center}
\label{quartic-max}
\end{table}%

We then consider the polynomial with $c_1\neq 0$, for which the phase $\Phi(\theta,\vec{c})$ in equation (\ref{phase}) is nontrivial. Evaluating the masses (period integrals), we find $m_1\neq m_3$ and $m_{12}\neq m_{23}$.  In Table \ref{r0-z4-z2-z-1} and Table \ref{quartic-z-dr}, we list $r_0(\vec{c})$ and $r(\vec{c})$, respectively, as defined in  (\ref{rdefinitions}). In Table \ref{quartic-z-dr-max}, we show the $\frac{6}{\pi}F$ and $\frac{6}{\pi}A_{\rm free}$ for the polynomial $p(z, \vec{c})=z^4+c_2 z-\frac{z}{200}-1$ with several $c_2$.
\begin{table}[H]
\caption{\footnotesize  $r_0(\vec{c}), r_0(\vec{c}^{R_1}), r_0(\vec{c}^{R_2})$ and  $r_0(\vec{c}^{R_3})$ for $p(z,\vec{c})=z^4+c_2z^2-\frac{z}{200}-1$. $r(\vec{c}^{R_3})$ is obtained by $r_0(\vec{c}^{R_3})=r_0(\vec{c}^{R_1})^\ast$.}
\begin{center}
\begin{tabular}{c|c|c|c}
$c_2$&$r_0(\vec{c})$ & $r_0(\vec{c}^{R_1})$& $r_0(\vec{c}^{R_2})$ \\\hline
$1$ &$2.580322548331587` $&$5.049378533767665`-0.002318241884034844`  i$ &$2.5849590320996567` $ \\\hline
$\frac{1}{2}$ &$2.972598628551023`$ & $4.181167930139899`-0.0012249121993099799` i$&$2.9750484529496433` $ \\\hline
$\frac{1}{100}$ &$3.4841068384904066` $  & $3.5080872072271383`-0.000024977324456942895`  i$&$3.4841567931393205` $\\\hline
\end{tabular}
\end{center}
\label{r0-z4-z2-z-1}
\end{table}%
\begin{table}[H]
\caption{\footnotesize  $r(\vec{c}), r(\vec{c}^{R_1}), r(\vec{c}^{R_2})$ and  $r(\vec{c}^{R_3})$ for the polynomial $p(z,\vec{c})=z^4+c_2z^2-\frac{z}{200}-1$. $r(\vec{c}^{R_3})$ is obtained by $r(\vec{c}^{R_1})=r(\vec{c}^{R_1})^\ast$}
\begin{center}
\begin{tabular}{c|c|c|c}
$c_2$&$r(\vec{c})$ & $r(\vec{c}^{R_1})$& $r(\vec{c}^{R_2})$\\\hline
$1$ & $2.5881765299655615`$ & $ 5.049378533767665` - 0.010172223518009327` i$&$2.5771050504656823`$ \\\hline
$\frac{1}{2}$ & $2.9804526101849973`$ & $4.181167930139899` - 0.009078893833284463`i$&$2.967194471315669`$ \\\hline
$\frac{1}{100}$ &$3.491960820124381`$  & $3.5080872072271383` - 0.007878958958431426`i $&$3.476302811505346`$\\\hline
\end{tabular}
\end{center}
\label{quartic-z-dr}
\end{table}%
\begin{table}[H]
\caption{\footnotesize  Numeric check in maximal chamber: $p(z,\vec{c})=z^4+c_2z^2-\frac{z}{200}-1$. The NLIEs are solved by Fourier discretisation with cutoff $(-12,12)$ and $2^{12}$ points.  The parameter $\epsilon$ in eq.(3.15) is fixed to be $1/2^4$. The TBA equations are solved by $2^{14}$ points and cutoff $(-20,20)$.}
\begin{center}
\begin{tabular}{c|c|c|c}
$c_2$&$\frac{6}{\pi}F$ & $\frac{6}{\pi}A_{\rm free}$& $F/A_{\rm free}$\\\hline
$1$ & $0.037104841463217055`$ & $0.03710484211496269` $&$0.9999999824350246`$ \\\hline
$\frac{1}{2}$ & $0.04910375701712826`$ & $0.04910376260115915` $&$0.9999998862809978`$ \\\hline
$\frac{1}{100}$ &$0.05536470669758174`$  & $0.055364704965062565`$&$1.0000000312928459`$\\\hline
\end{tabular}
\end{center}
\label{quartic-z-dr-max}
\end{table}%

We then fix $c_2$ to be $1$, and vary $c_1$. In Table \ref{quartic-z-lg-r0} and Table \ref{quartic-z-lg-dr}, we list the $r_0(\vec{c})$ and $r(\vec{c})$ respectively for this case. In Table \ref{quartic-z-lg-dr-max}, we show the $\frac{6}{\pi}F$ and $\frac{6}{\pi}A_{\rm free}$ for the polynomial $p(z, \vec{c})=z^4+z^2+c_1z-1$.
\begin{table}[H]
\caption{\footnotesize  $r_0(\vec{c}), r_0(\vec{c}^{R_1}), r_0(\vec{c}^{R_2})$ and $r_0(\vec{c}^{R_3})$ for the polynomial $p(z,\vec{c})=z^4+z^2+c_1z-1$. $r_0(\vec{c}^{R_3})$ is obtained by $r_0(\vec{c}^{R_1})=r_0(\vec{c}^{R_1})^\ast$.}
\begin{center}
\begin{tabular}{c|c|c|c}
$c_1$&$r_0(\vec{c})$ & $r_0(\vec{c}^{R_1})$& $r_0(\vec{c}^{R_2})$\\\hline
$-\frac{1}{2}$ & $2.3933954392247605` $ &$5.037001372706346` - 0.23424307835549674` i$&$2.8618815959357526` $ \\\hline
$-\frac{1}{4}$ & $2.4777183763893467` $&$5.046201443952038` - 0.11622181204658544`i$&$2.7101620004825175` $ \\\hline
$-\frac{1}{50}$ &$2.5734354362619722` $&$5.049359181799129` - 0.009273172229004922` i$&$2.591981780719982` $\\\hline
\end{tabular}
\end{center}
\label{quartic-z-lg-r0}
\end{table}%
\begin{table}[H]
\caption{\footnotesize $r(\vec{c}), r(\vec{c}^{R_1}), r(\vec{c}^{R_2})$ and $r(\vec{c}^{R_3})$ for the polynomial $p(z,\vec{c})=z^4+z^2+c_1z-1$. $r(\vec{c}^{R_3})$ is obtained by $r(\vec{c}^{R_1})=r(\vec{c}^{R_1})^\ast$.}
\begin{center}
\begin{tabular}{c|c|c|c}
$c_1$&$r(\vec{c})$ & $r(\vec{c}^{R_1})$& $r(\vec{c}^{R_2})$\\\hline
$-\frac{1}{2}$ & $3.178793602622209`$ & $5.037001372706346` - 1.019641241752945` i$&$2.0764834325383044`$ \\\hline
$-\frac{1}{4}$ & $2.870417458088071`$ & $5.046201443952038` - 0.5089208937453096` i$&$2.3174629187837934`$ \\\hline
$-\frac{1}{50}$ &$2.60485136279787`$  & $ 5.049359181799129` - 0.040689098764902856`i$&$2.5605658541840843`$\\\hline
\end{tabular}
\end{center}
\label{quartic-z-lg-dr}
\end{table}%
\begin{table}[H]
\caption{\footnotesize  Numeric check in maximal chamber: $p(z,\vec{c})=z^4+z^2+c_1 z-1$. The NLIEs are solved by Fourier discretisation with cutoff $(-12,12)$ and $2^{10}$ points.  The parameter $\epsilon$ in eq.(3.15) is fixed to be $1/2$. The TBA equations are solved by $2^{14}$ points and cutoff $(-20,20)$.}
\begin{center}
\begin{tabular}{c|c|c|c}
$c_1$&$\frac{6}{\pi}F$ & $\frac{6}{\pi}A_{\rm free}$& $F/A_{\rm free}$\\\hline
$-\frac{1}{2}$ & $0.03704085160351035`$ & $0.03704085160351035`$&$1.000000000000000$ \\\hline
$-\frac{1}{4}$ & $0.03716372850655718`$ & $0.03716372633241627`  $&$1.000000058501693`$ \\\hline
$-\frac{1}{50}$ &$0.03710536684451506`$  & $0.037105361745906645` $&$1.0000001374089398`$\\\hline
\end{tabular}
\end{center}
\label{quartic-z-lg-dr-max}
\end{table}%

In a similar way, we have tested our NLIEs (3.15) for the polynomial $p(z, \vec{c})=z^3+c_1z-1$ against the TBA equations in the maximal chamber\footnote{The TBA equations can be found in equation (3.80) of \cite {IMS}.}. In Table \ref{cubic-max}, we present $\frac{6}{\pi}F$ and $\frac{6}{\pi}A_{\rm free}$ for several values of $c_1$.
\begin{table}[H]
\caption{\footnotesize  $\frac{6}{\pi}F$  and $\frac{6}{\pi}A_{\rm free}$  for $p(z, \vec{c})=z^3+c_1 z-1$. The NLIEs are solved by Fourier discretization with cutoff $(-12,12)$ and $2^{12}$ points.  The parameter $\epsilon$ in eq.(3.15) is fixed to be $1/2^4$. The TBA equations are solved by $2^{14}$ points and cutoff $(-20,20)$.}
\begin{center}
\begin{tabular}{c|c|c|c}
$c_1$ &$\frac{6}{\pi}F$ & $\frac{6}{\pi}A_{\rm free}$ & $F/A_{\rm free}$\\ \hline
$-\frac{1}{2}$ & $0.027577390204736915` $ & $0.027585003289666222` $ &$0.9997240136298206`$\\\hline
$-\frac{1}{10}$ & $0.017482596263586746`$ & $0.017482865359751576` $&$0.9999846080056505`$ \\\hline
$-\frac{1}{100}$ & $0.01719846067124951`$ & $0.017198793671935574` $&$0.9999806381370412`$ \\\hline
\end{tabular}
\end{center}
\label{cubic-max}
\end{table}%

Our numerical results show that
\be
F=A_{\rm free} \, ,
\ee
at least in maximal chamber. In other chambers, we need to consider the wall crossing of NLIEs and TBA equations, which will be pursued in a future publication. At the end of section, it is worth to note that our NLIEs are much simpler than the TBA equations, especially when the number of cusps of Wilson loop is large. For the lightlike Wilson loop with $4N+4$ cusps, one needs $2N(2N-1)/2$ TBA equations to compute the free energy in maximal chamber, while the number of NLIEs is only $2N$. Moreover, one needs to perform $(N-1)(2N-1)$ times wall crossing to derive the TBA equations in the maximal chamber, which is quite complicated. Therefore, it is much convenient to use the NLIEs instead of TBA equation, especially in maximal chamber.

\section{Conformal limit}
\setcounter{equation}{0}
\label{conformal-case}

Consider the NLIEs in the limit
\be
\label{clim}
|c_0| \rightarrow 0 \, , \quad c_n =|c_0|^{\frac{2N-n}{2N}} c_n^{cft} \rightarrow 0 \, , \quad \theta =\theta ^{cft}-
\frac{1+N}{2N}\ln |c_0| \rightarrow +\infty \, ,
\ee
with $c_n^{cft}$, $\theta ^{cft}$ finite. Then, we have that
\be
w_0(\vec{c})e^{\theta} \rightarrow w_0(\vec{c}^{cft})e^{\theta ^{cft}} \, , \quad \bar w_0(\vec{c})e^{-\theta} \rightarrow 0 \, , \quad B_{-1}(\vec{c})e^{\theta} \rightarrow  B_{-1}(\vec{c}^{cft})e^{\theta ^{cft}}
 \, , \quad \bar B_{-1}(\vec{c})e^{-\theta} \rightarrow  0 \, .
\ee
The linear problem in the variable $z$, after the scaling
\be
z=xe^{-\frac{\theta}{1+N}} \, ,
\ee
becomes
\be
\left [ -\frac{d^2}{dx^2}+\frac{l(l+1)}{x^2}+\sum _{n=1}^{2N} c_n^{cft}e^{\frac{2N-n}{1+N}\theta ^{cft}}x^n \right ]
\psi (x)= e^{\frac{2N}{1+N}\theta ^{cft}}\psi (x) \, ,
\ee
which is a Schr\"{o}dinger problem with energy $E=e^{\frac{2N}{1+N}\theta ^{cft}}$ and potential
\be
\label{potential}
V(x)=\sum _{n=1}^{2N} c_n^{cft}e^{\frac{2N-n}{1+N}\theta ^{cft}}x^n \, .
\ee
Interestingly, the limiting $\hat \Omega$ symmetry acts as
\be
\hat \Omega _{lim}: \quad x\rightarrow x e^{\frac{i\pi}{1+N}} \, , \quad
c_n^{cft} \rightarrow c_n^{cft} e^{i\pi\frac{2N-n}{1+N}}
\ee
and the limiting $\hat \Pi$ symmetry as
\be
\hat \Pi _{lim}: \quad x\rightarrow x e^{-\frac{i\pi}{1+N}} \, , \quad
c_n^{cft} \rightarrow c_n^{cft} e^{-i\pi\frac{2N-n}{1+N}} \, .
\ee
Therefore in the ODE limit $\hat \Omega _{lim}=(\hat \Pi _{lim})^{-1}$ and then one can use only one type of symmetry.

\medskip

In this limit, we make of course contact with known results. For instance, if $p(z, \vec{c})=z^{2N}+c_{N-1}z^{N-1}+c_0$, then $B_{-1}(\vec{c})=c_{N-1}/2$ and the extra driving term of the NLIEs, $-\frac{\pi}{N}B_{-1}(\vec{c}^{cft})e^{\theta ^{cft}}=\frac{-\pi c_{N-1}^{cft}}{2N}e^{\theta ^{cft}}$, equals $-\pi/2N$ times the coefficient of the term $x^{N-1}$ in $V(x)$, in agreement with \cite{SUZ00} and \cite{DDT01}.

To complete our comparison between our NLIEs in conformal limit and the NLIEs found in \cite{SUZ00} and \cite{DDT01}, we report here some numerical results. We consider as an example the polynomial $p(z, \vec{c})=z^4+|c_0|^{3/4}c_1^{cft}z-|c_0|$ with a small negative $c_0$. Under the limit (\ref{clim}), $w_0(\vec{c})e^\theta$ becomes
\be
w_0(\vec{c})e^\theta \to -\int_0^\infty d\tilde{z} \Big[\sqrt{\tilde{z}^4+e^{\frac{3\pi i}{4}}\tilde{c}_1^{cft}\tilde{z}+1}-q_2(\tilde{z})\Big]e^{\theta^{cft}} \, ,
\ee
where $z=|c_0|^{1/4}\tilde{z}$ and $q_2(\tilde{z})$ is the related polynomial which makes the integration convergent.
In Table \ref{DDT-x4-r0} and \ref{DDT-x4-r}, we list $r_0({c}^{cft}_1)$ and $r({c}^{cft}_1)$
\begin{table}[H]
\caption{\footnotesize  $r_0((c^{cft}_1)^{R^n})$ in the conformal limit. $r_0((c^{cft}_1)^{R^3})$ is obtained by conjugating $r_0(({c}^{cft}_1)^{R^1})$)}
\begin{center}
\begin{tabular}{c|c|c|c}
$c_1^{cft}$&$r_0(c_1^{cft})$ & $r_0((c_1^{cft})^{R_1})$& $r_0((c_1^{cft})^{R_2})$\\\hline
$-\frac{1}{100}$ & $3.4960916735302283`$ & $3.4960617618285434`   $&$3.496091757138591`$ \\\hline
$-\frac{1}{10}$ & $3.4975335418072313`  $ & $3.4945799178619184` - 0.00004172438919569643` i$&$3.4976169905856227` $ \\\hline
\end{tabular}
\end{center}
\label{DDT-x4-r0}
\end{table}%
\begin{table}[H]
\caption{\footnotesize $r(({c}^{cft}_1)^{R^n})$ in the conformal limit. $r(({c}^{cft}_1)^{R^3})$ is obtained by conjugating $r(({c}^{cft}_1)^{R^1})$)}
\begin{center}
\begin{tabular}{c|c|c|c}
$c_1^{cft}$&$r(c_1^{cft})$ & $r((c_1^{cft})^{R_1})$& $r((c_1^{cft})^{R_2})$\\\hline
$-\frac{1}{100}$ & $3.511799636798177`$ & $3.4960617618285434` - 0.015707963267948967`i $&$3.4803837938706423`$ \\\hline
$-\frac{1}{10}$ & $3.654613174486721`$ & $3.4945799178619184` - 0.15712135706868535` i$&$3.340537357906133`$ \\\hline\end{tabular}
\end{center}
\label{DDT-x4-r}
\end{table}%
Moreover, the driving terms in NLIEs (\ref{nlies}) become $\frac{1}{2}|r(({c}^{cft}_1)^{R^n})|e^{\theta^{cft}}$ in the conformal limit. With these driving terms we evaluate (\ref{nlies}) in the conformal limit numerically.

\begin{figure}[H]
\begin{center}
\begin{tabular}{cc}
\resizebox{85mm}{!}{\includegraphics{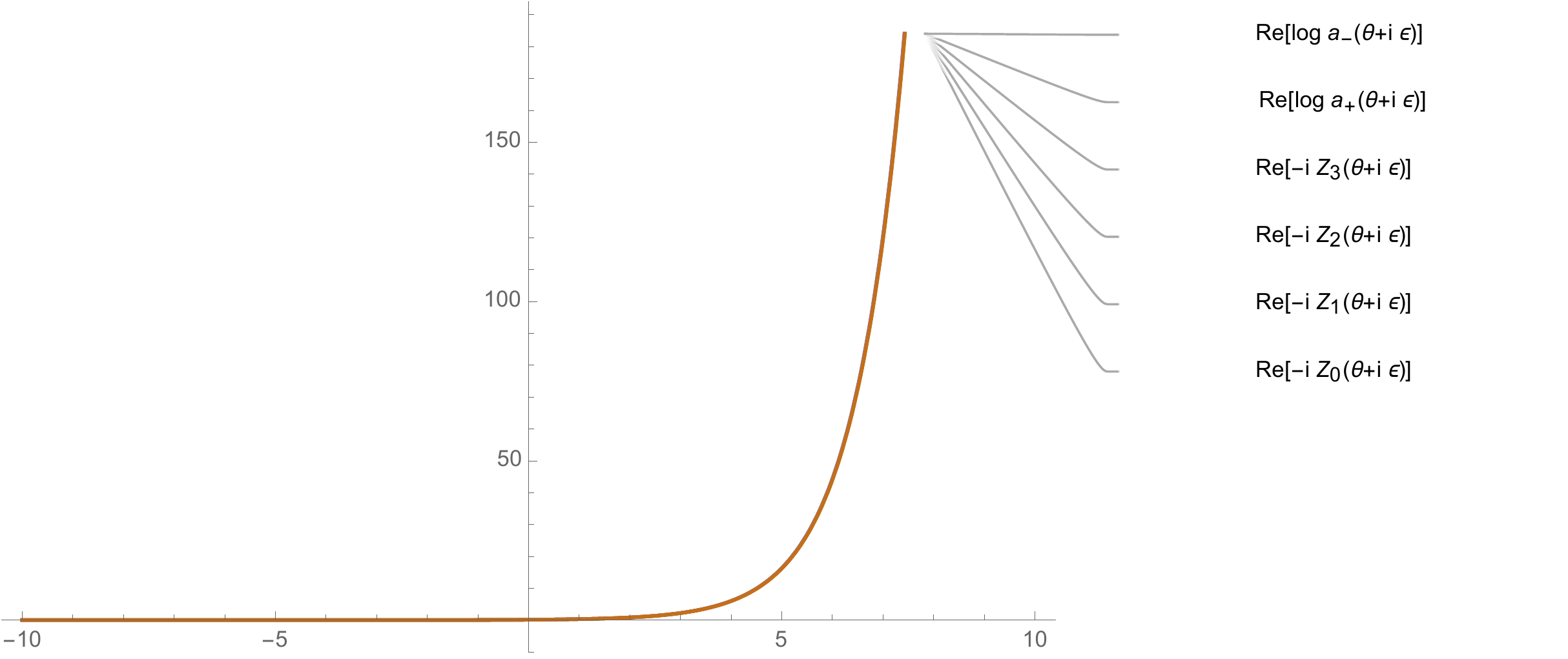}}
\hspace{0mm}
&
\resizebox{100mm}{!}{\includegraphics{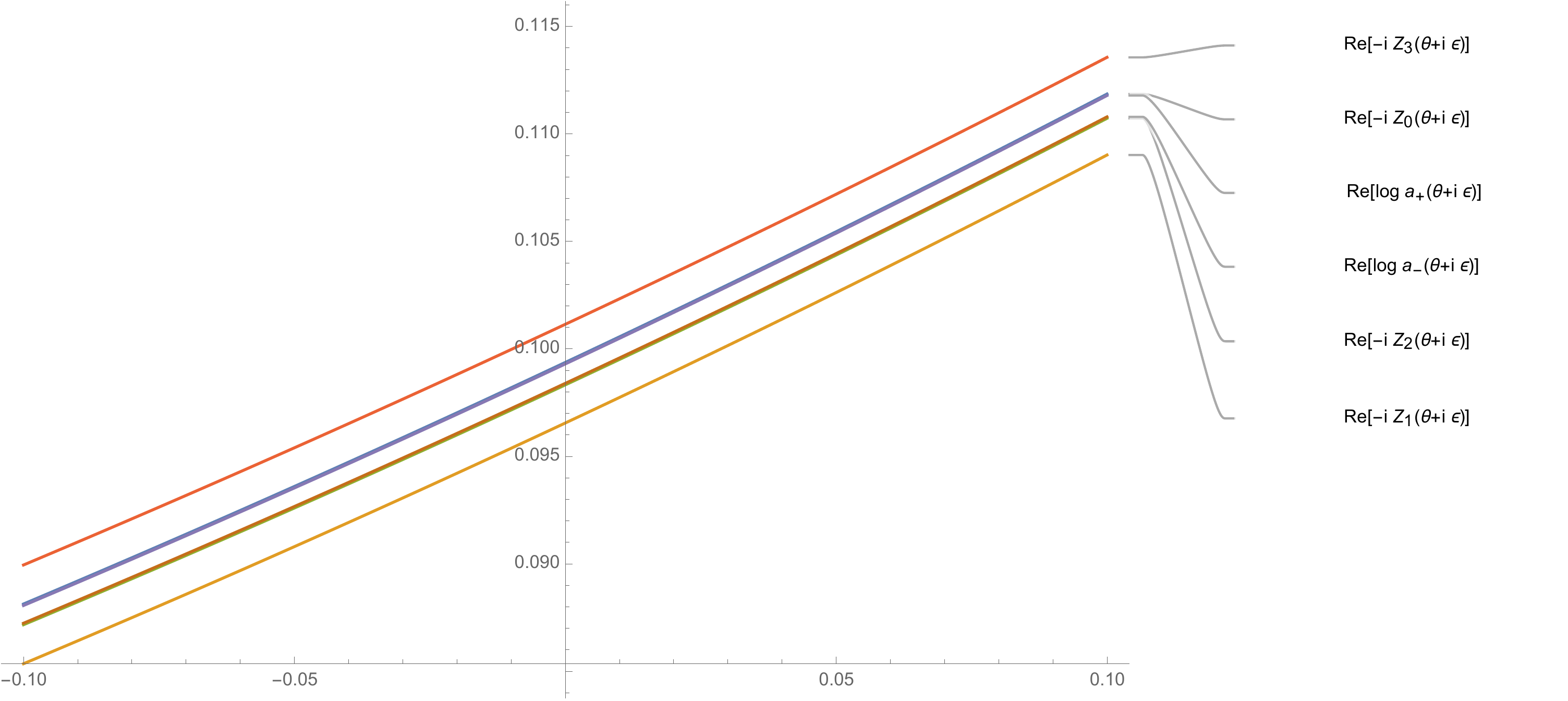}}
\end{tabular}
\end{center}
  \caption{The real parts of the functions $-iZ(\theta^{cft}-i\phi_n,\vec{c}^{R^n})$ and $\ln a_\pm(\theta^{cft})$ with $c_1^{cft}=-\frac{1}{100}$.}
\label{fig:r100Za}
\end{figure}

\begin{figure}[H]
\begin{center}
\begin{tabular}{cc}
\resizebox{85mm}{!}{\includegraphics{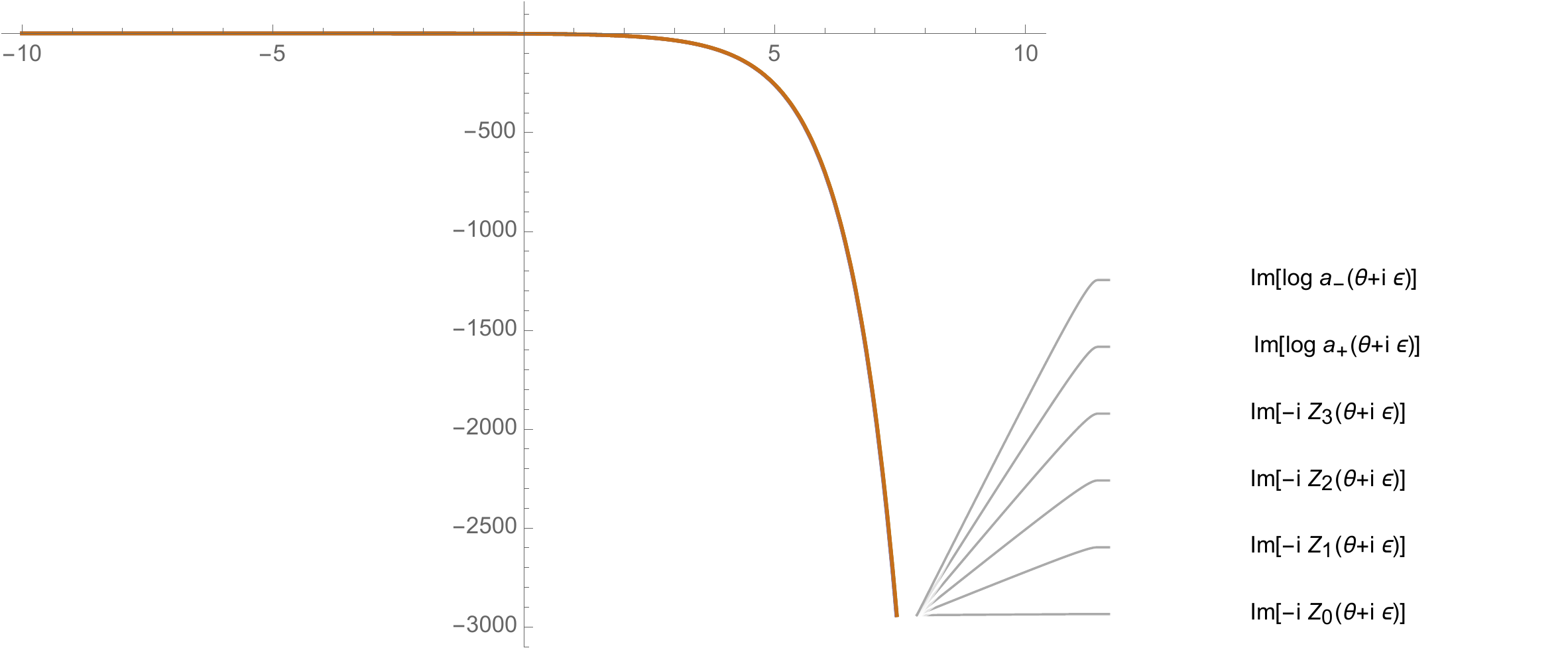}}
\hspace{0mm}
&
\resizebox{100mm}{!}{\includegraphics{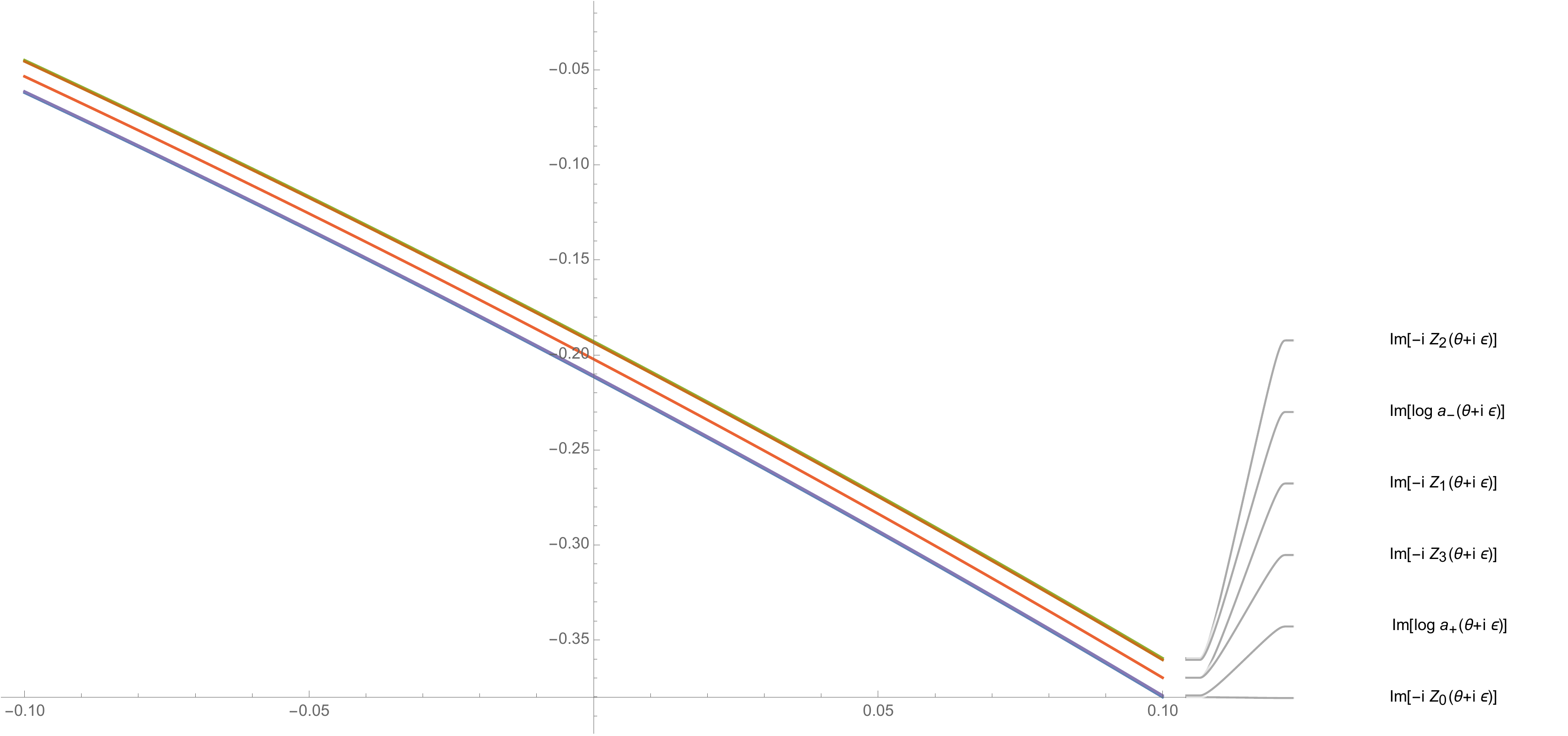}}
\end{tabular}
\end{center}
  \caption{The imaginary part of the functions $-iZ(\theta^{cft}-i\phi_n,\vec{c}^{R^n})$ and $\ln a^\pm(\theta^{cft})$ with $c_1^{cft}=-\frac{1}{100}$.}
\label{fig:I100Za}
\end{figure}

On the other hand, starting from $p(z)=z^4+|c_0|^{3/4}c_1^{cft}z-|c_0|$, the potential (\ref{potential}) becomes $V(x)=x^4+c_1^{cft}e^{\theta}x$, from which one finds the NLIEs \cite{SUZ00} and \cite{DDT01}:
\be
\begin{aligned}
\label{nlies-SDDT}
\ln a_{\pm}(\theta^{cft})&=\frac{\pi i}{2}(2l+1\pm\frac{\alpha}{2})-ib_{0}e^{\theta^{cft}}\\
&\quad+\int_{{\cal C}_{1}}d\theta^{\prime}K_{1}(\theta^{cft}-\theta^{\prime})\ln\big(1+a_{\pm}(\theta^{\prime})\big)-\int_{{\cal C}_{2}}d\theta^{\prime}K_{1}(\theta^{cft}-\theta^{\prime})\ln\big(1+\frac{1}{a_{\pm}(\theta^{\prime})}\big)\\
&\quad+\int_{{\cal C}_{1}}d\theta^{\prime}K_{2}(\theta^{cft}-\theta^{\prime})\ln\big(1+a_{\mp}(\theta^{\prime})\big)-\int_{{\cal C}_{2}}d\theta^{\prime}K_{2}(\theta^{cft}-\theta^{\prime})\ln\big(1+\frac{1}{a_{\mp}(\theta^{\prime})}\big),
\end{aligned}
\ee
where $\alpha=c_1^{cft}e^\theta$ and the kernels $K_1$ and $K_2$ are given by
\be
\begin{aligned}
K_{1}(\theta^{cft})&=-\big(\frac{1}{4\pi\cosh\theta^{cft}}-\frac{\theta^{cft}}{2\pi^{2}\cosh\theta^{cft}\sinh\theta^{cft}}\big),\\ K_{2}(\theta^{cft})&=-\big(\frac{1}{4\pi\cosh\theta^{cft}}+\frac{\theta^{cft}}{2\pi^{2}\cosh\theta^{cft}\sinh\theta^{cft}}\big).
\end{aligned}
\ee
${\cal C}_1$ and ${\cal C}_2$ are integration contours just below and just above the real $\theta^{cft}$ axis with a distance $\epsilon$.
We introduce the related free energy obtained from these NLIEs by
\be
\begin{aligned}
{\cal F}&=\frac{i}{\pi}\int d\theta^{cft}\Big((b_{0}+\frac{\pi}{4}c_{1}^{cft})e^{\theta^{cft}-i\epsilon}\ln\big(1+a_{+}(\theta^{cft}-i\epsilon)\big)+(b_{0}-\frac{\pi}{4}c_{1}^{cft})e^{\theta^{cft}-i\epsilon}\ln\big(1+a_{-}(\theta^{cft}-i\epsilon)\big)\\
&-(b_{0}+\frac{\pi}{4}c_{1}^{cft})e^{\theta^{cft}+i\epsilon}\ln\big(1+\frac{1}{a_{+}(\theta^{cft}+i\epsilon)}\big)-(b_{0}-\frac{\pi}{4}c_{1}^{cft})e^{\theta^{cft}+i\epsilon}\ln\big(1+\frac{1}{a_{-}(\theta^{cft}+i\epsilon)}\Big).
\end{aligned}
\ee
Then, our free energy in conformal limit $F^{cft}$ and ${\cal F}$ lead to the same result
\be
F^{cft}=\frac{\pi}{6}={\cal F} \, ,
\ee
when $l=0$. In order to understand the precise relationship between our NLIEs and the ones of (\ref{nlies-SDDT}), we solve both numerically. We evaluate the two type of NLIEs by Fourier discretization with a cutoff $(-12, 12)$ and $2^{10}$ points. The parameter $\epsilon$ related with integration contour is fixed to be $1/2^4$.
In Fig.\ref{fig:r100Za} and Fig.\ref{fig:I100Za}, we plot the real and imaginary parts of the functions $-iZ(\theta-i\phi_n,\vec{c}^{R^n})$ and $\ln a_\pm(\theta)$, respectively, with fixed $c_1^{cft}=-\frac{1}{100}$ and $l=0$.

\begin{figure}[H]
\begin{center}
\begin{tabular}{cc}
\resizebox{85mm}{!}{\includegraphics{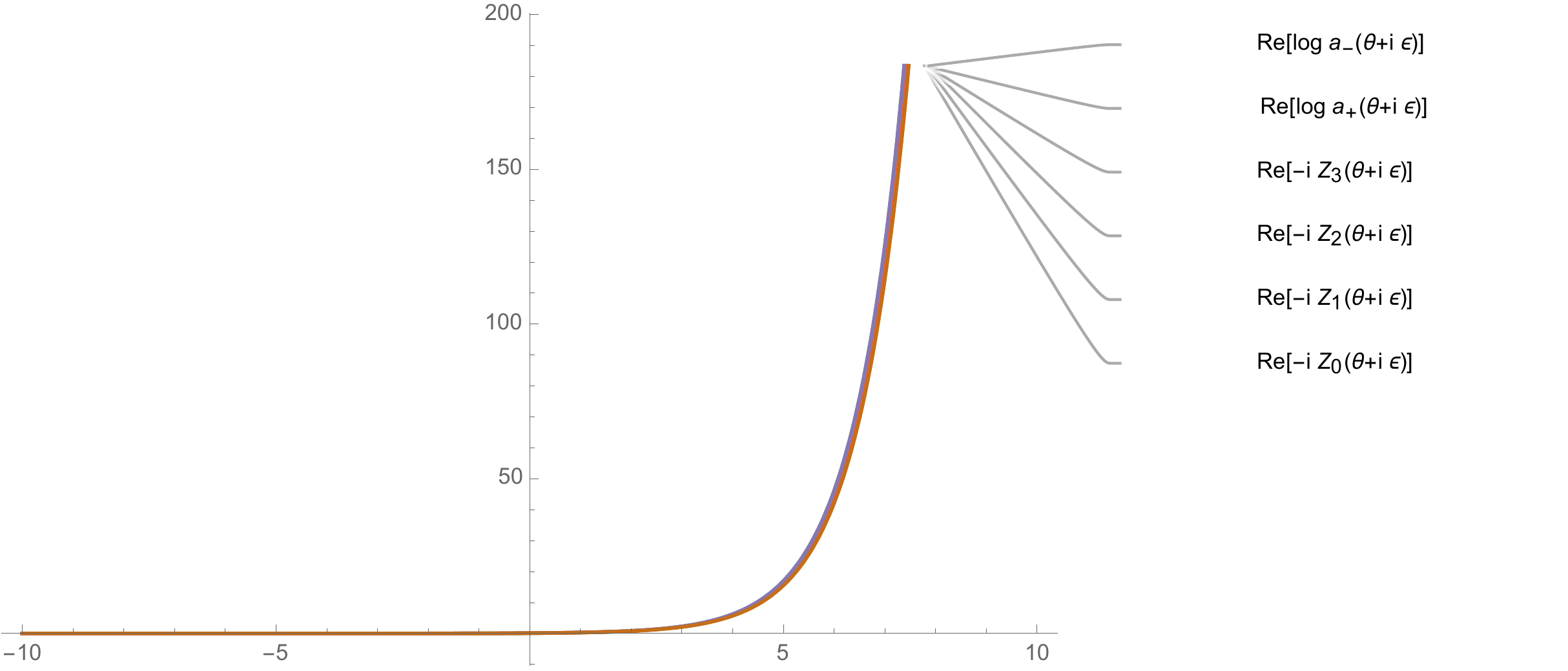}}
\hspace{0mm}
&
\resizebox{100mm}{!}{\includegraphics{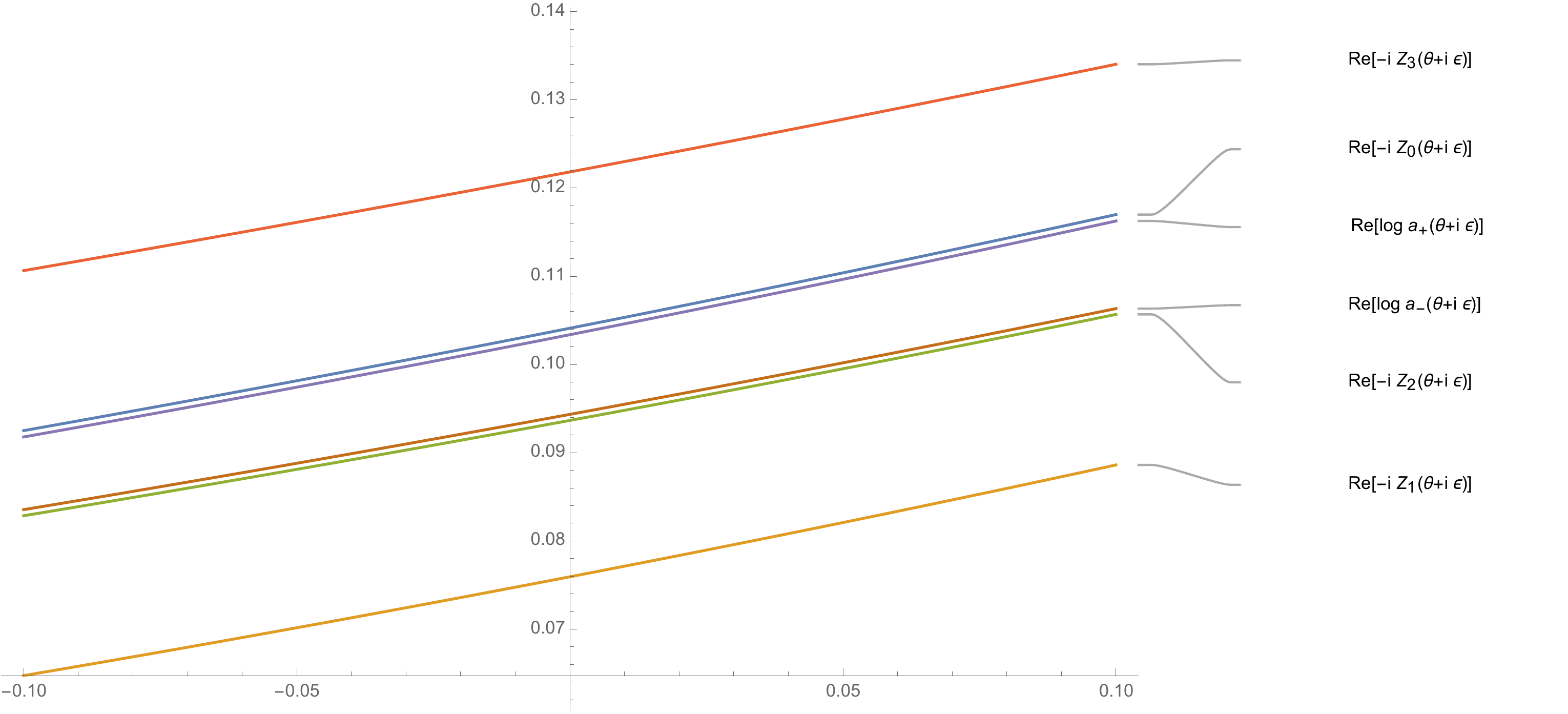}}
\end{tabular}
\end{center}
  \caption{The real parts of the functions $-iZ(\theta^{cft}-i\phi_n,\vec{c}^{R^n})$ and $\ln a_\pm(\theta^{cft})$ with $c_1^{cft}=-\frac{1}{10}$ and $l=0$.}
\label{fig:r10Za}
\end{figure}

\begin{figure}[H]
\begin{center}
\begin{tabular}{cc}
\resizebox{85mm}{!}{\includegraphics{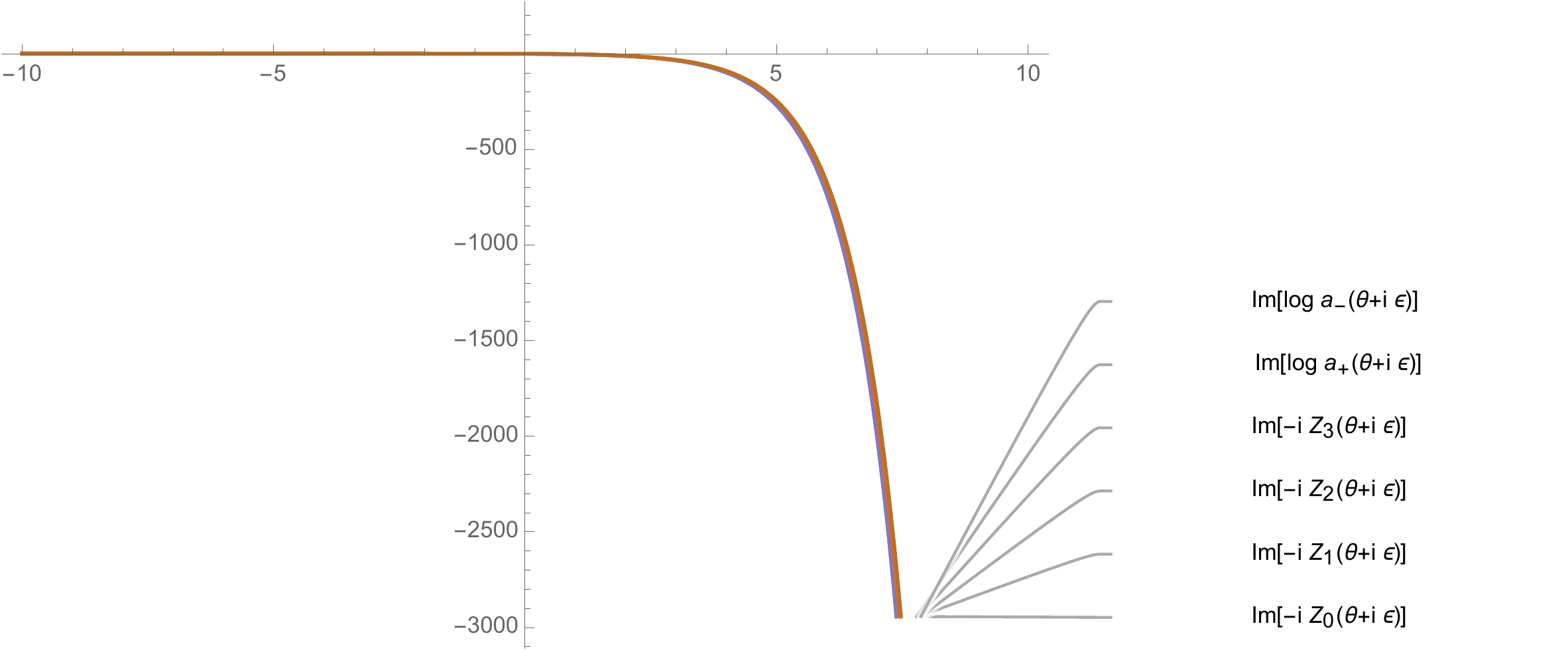}}
\hspace{0mm}
&
\resizebox{100mm}{!}{\includegraphics{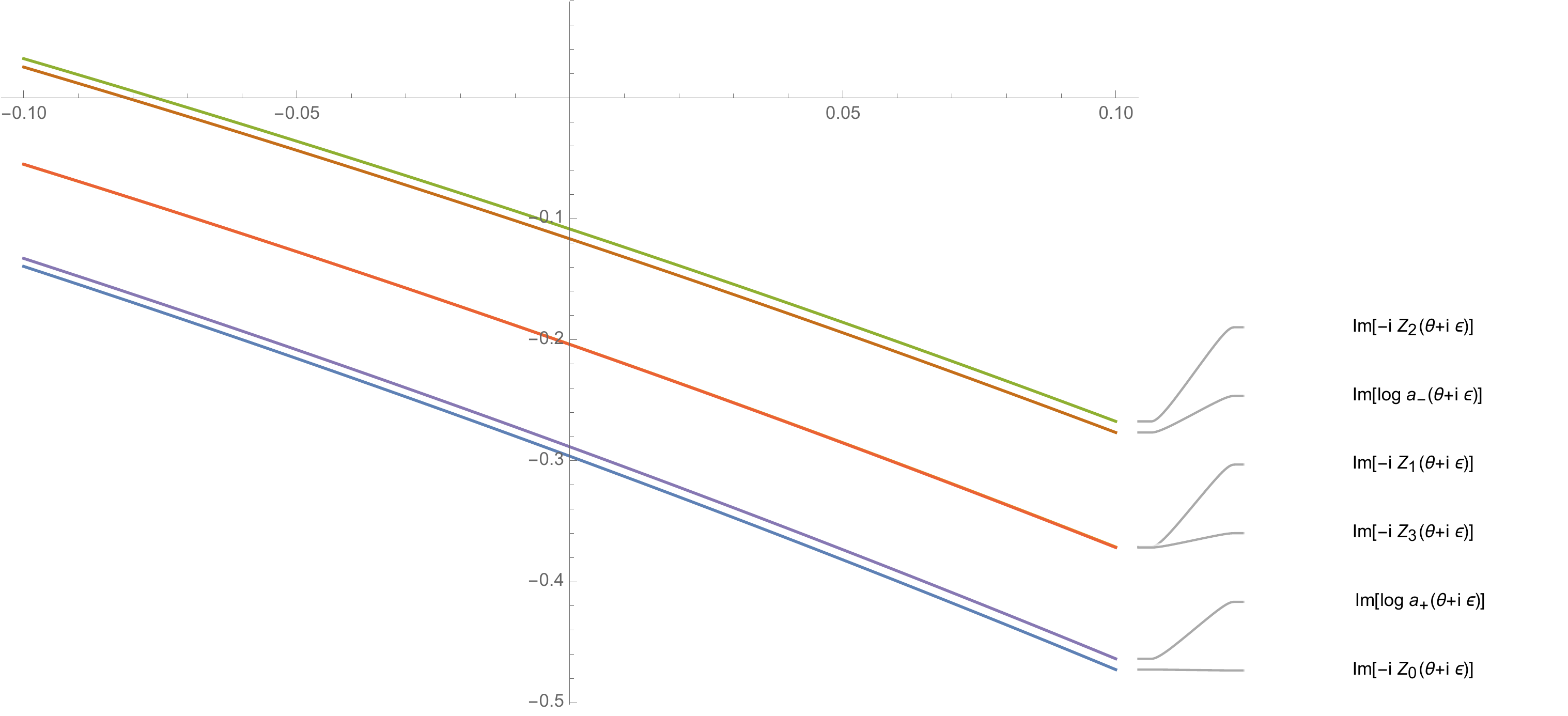}}
\end{tabular}
\end{center}
  \caption{The imaginary part of the functions $-iZ(\theta^{cft}-i\phi_n,\vec{c}^{R^n})$ and $\ln a_\pm(\theta^{cft})$ with $c_1^{cft}=-\frac{1}{10}$ and $l=0$.}
\label{fig:I10Za}
\end{figure}

In Fig.\ref{fig:r10Za} and Fig.\ref{fig:I10Za}, we plot the real and imaginary parts of the functions $-iZ(\theta-i\phi_n,\vec{c}^{R^n})$ and $\ln a_\pm(\theta)$, respectively, with fixed $c_1^{cft}=-\frac{1}{10}$ and $l=0$.

These numeric results (and many other examples) suggest
\be
\begin{aligned}
\label{Z-a}
-iZ(\theta^{cft}-i\phi_0,\vec{c}^{cft})=\ln a_+(\theta^{cft}),\quad
-iZ(\theta^{cft}-i\phi_2,(\vec{c}^{cft})^{R^2})=\ln a_-(\theta).
\end{aligned}
\ee
From the numeric solution, we also find
\be
\begin{aligned}
\label{Z-Z}
Z(\theta^{cft}-i\phi_0,\vec{c}^{cft})+Z(\theta^{cft}-i\phi_2,(\vec{c}^{cft})^{R^2})&=Z(\theta-i\phi_1,(\vec{c}^{cft})^{R^1})+Z(\theta-i\phi_3,(\vec{c}^{cft})^{R^3}),\\
{\rm Re}(Z(\theta^{cft}-i\phi_1,(\vec{c}^{cft})^{R^1})&={\rm Re}(Z(\theta^{cft}-i\phi_3,(\vec{c}^{cft})^{R^3}) \, .
\end{aligned}
\ee
Moreover, we also find numerically the same relations (\ref{Z-a}) and (\ref{Z-Z}) for various values of $l$ between $-\frac{1}{2}$ and $+\frac{1}{2}$. Then, we can very reasonably conclude that the conformal limit of our NLIEs (\ref{nlies}) reproduce the results of \cite{SUZ00} and \cite{DDT01}.

\section{Conclusions}

We have constructed the main functional relations, $QQ$- and $TQ$- systems, underlying the (integrable) computations of the gluon scattering amplitudes in (planar) $\mathcal{N} = 4$ SYM at strong coupling. Kinematical data of scattered gluons are encapsulated in complex quantities, the moduli,
which appear 'dynamically' in the functional relations. In addition, we have provided a system of coupled NLIEs which renders effective the analytical and numerical explicit computation of these amplitudes. We tested our equations in the case of regular polygons (conformal limit) and the octagon ('harmonic oscillator' limit) by means of explicit analytic computations. In addition we provided various numerical tests for polynomials corresponding to polygons with ten and twelve sides.

For the future, the principal aim is that our disentanglement of the integrability structure of the problem could shed light on the string quantisation, which, in the end, should reduce to a suitable deformation of the classical model (\ref{cl-sinh}, \ref{ass-lin-prob}, \ref{D}, \ref{barD}) or the derived functional equations or the Bethe Ansatz equations (\ref{BAEqs}).

Then, it would be of interest to connect the linear problems with the strong coupling limit of an OPE series \cite {BSV,FPR}, possibly its restriction to $AdS_3$. On the other way around, it is very useful to generalise our method, based on $SU(2)$, to the full $AdS_5$ (at strong coupling), the $SU(4)$ Hitchin system indeed.

In addition, the formalism discussed in this paper could be applied also to Wilson loops in Euclidean $AdS_3$, which problem becomes, after Pohlmeyer reduction, the modified cosh-Gordon equation \cite {KRU}.

\medskip

An important final remark concerns the solution of classical equation (\ref {cl-sinh}). In a forthcoming publication \cite {GLM-FR} we show that it is obtained as a solution of an equation with the form of Gelfand-Levitan-Marchenko  \cite {GLM} for which the inhomogeneous term (scattering data) is provided by $T$, the eigenvalue of the transfer matrix. This form of the equation is particularly important as it admits the solution to be written in terms of Fredholm determinants ({\it i.e.} tau functions)  
\be
\eta= \hat \eta + \frac{1}{4}\ln P\bar P \, , \quad \hat \eta =\ln \frac{\textrm{det}(1+K)}{\textrm{det}(1-K)} \,  
 \label{etasol2}
 \ee
and then, for instance, as a series expansion
\be
\hat \eta =\sum_{n=1}^{\infty} \hat \eta _{2n-1}  , \quad \hat \eta _{2n-1}=\frac{2}{2n-1}
\int  \prod _{i=1}^{2n-1} \frac{d\theta_i}{4\pi} T \left (\theta _i + \frac{i\pi}{2} +\frac{i\pi}{2N}, \vec{c} \right ) \frac{e^{-2we^{\theta _i}-2\bar w e^{-\theta _i}}}{\cosh \frac{\theta _i -\theta _{i+1}}{2}} \label{etasol2} \, .
\ee
$\eta $ is real due to the property $\bar T(\theta , \vec{c})=T(-\bar \theta , \vec{c})$, which comes from (\ref {qcompl}, \ref {trmat}).
And this is in fact the explicit solution of the initial problem, for $l=0$ the string problem\footnote{In fact, from it one can reconstruct the classical string world-sheet in $AdS_3$ by solving the two auxiliary linear problems at specific values of $\theta$.}, and we may think that a similar solution should exist for the more general $AdS_5$ case in terms of classical tau functions for the $SU(4)$ Hitchin system.

\vspace{1truecm}

{\bf Acknowledgements}
D.F. thanks R. Tateo and D. Masoero for stimulating discussions. This work has been partially supported by the grants: GAST (INFN), the EC Network Gatis and the MIUR-PRIN contract 2017CC72MK$\_$003. The work of H.S. has been supported by the grant ``Exact Results in Gauge and String Theories'' from the Knut and Alice Wallenberg foundation. H.S. would like to thank INFN and University of Bologna for the warm hospitality.

\end{document}